\documentclass[12pt, draftclsnofoot, onecolumn]{IEEEtran}

\usepackage[cmex10]{amsmath}
\usepackage{verbatim,cite,listings,subfigure,times,setspace,psfrag,color}
\usepackage[ruled,vlined]{algorithm2e}
\usepackage[hyphens]{url}
\usepackage[T1]{fontenc}
\usepackage{textcomp}
\usepackage[pdftex]{graphicx}

\begin{document}

\title{SecureFind: Secure and Privacy-Preserving Object Finding via Mobile Crowdsourcing
\thanks{J. Sun, X. Jin and Y. Zhang are with the School of Electrical, Computer and Energy Engineering, Arizona State University, Tempe, AZ, 85287 (email: \{jcsun,xiaocong.jin,yczhang\}@asu.edu).}
\thanks{R. Zhang is with the Department of Electrical Engineering, University of Hawaii, Honolulu, HI, 96822 (email: ruizhang@hawaii.edu).}
}

\author{Jingchao~Sun,~\IEEEmembership{Student~Member,~IEEE,}
        Rui~Zhang,~\IEEEmembership{Member,~IEEE,}
        Xiaocong~Jin,~\IEEEmembership{Student~Member,~IEEE,}
        and~Yanchao~Zhang,~\IEEEmembership{Senior Member,~IEEE}% <-this % stops a space
%\IEEEcompsocitemizethanks{\IEEEcompsocthanksitem X. Jin, J. Sun and Y. Zhang are with the Department
%of Electrical Engineering, Arizona State University, Tempe, AZ, 85287 USA.\protect\\
%E-mail: xiaocong.jin@asu.edu.
%\IEEEcompsocthanksitem R. Zhang is with the Department of Electrical Engineering, University of Hawaii.}% <-this % stops an unwanted space
%%\thanks{Manuscript received April 19, 2005; revised December 27, 2012.}
}
%\markboth{IEEE TRANSACTIONS ON WIRELESS COMMUNICATIONS,~VOL.~XX, NO.~YY, XX~201X}%
%{Sun \MakeLowercase{\textit{et al.}}: SecureFind: Secure and Privacy-Preserving Object Finding via Mobile Crowdsourcing}

\maketitle

\begin{abstract}
The plummeting cost of Bluetooth tags and the ubiquity of mobile devices are revolutionizing the traditional lost-and-found service. This paper presents SecureFind, a secure and privacy-preserving object-finding system via mobile crowdsourcing. In SecureFind, a unique Bluetooth tag is attached to every valuable object, and the owner of a lost object submits an object-finding request to many mobile users via the SecureFind service provider. Each mobile user involved searches his vicinity for the lost object on behalf of the object owner who can infer the location of his lost object based on the responses from mobile users. SecureFind is designed to ensure strong object security such that only the object owner can discover the location of his lost object as well as offering strong location privacy to mobile users involved. The high efficacy and efficiency of SecureFind are confirmed by extensive simulations.
\end{abstract}


%%%%%%%%%%%%%%%%%%%%%%%%%%%%%%%%%%%%%%%%%%%%%%%%%%%%%%%%%%%%%%%
\section{Introduction}\label{sec:Introduction}
%%%%%%%%%%%%%%%%%%%%%%%%%%%%%%%%%%%%%%%%%%%%%%%%%%%%%%%%%%%%%%%
\IEEEPARstart{T}{he} loss and recovery of physical \emph{object}s is a significant issue around the world. Here an object can refer to anything valuable such as personal assets, children, elderly with dementia, and pets. For example, about 800,000 US children are reported lost each year \cite{LostChild},
113 cell phones are lost/stolen every minute in the US \cite{USPhoneTheft}, and 19,000 items are lost every year by New York subway and bus riders \cite{USPhoneTheft}. The predominant method for recovering lost objects is through a lost-and-found place, where lost objects are turned in and returned to their owners with proper identification. Many (if not most) lost objects, however, may not be found or turned in, and the object owner may not know which of the possibly many lost-and-found places he should resort to. The recovery rate for lost objects is thus very low. For instance, University of California Police reported only 19.3\% of lost items recovered \cite{USPhoneTheft}. In addition, the recovery latency of this traditional method may be too long to be useful. As an example, by the time a lost object is found and turned in to an airport office, the object owner may have departed to a different city or country.

The plummeting cost and ultra-low energy consumption of Bluetooth tags make them very promising to revolutionize the lost-and-found service. In contrast to RFID tags, Bluetooth tags can directly communicate with any mobile device with a Bluetooth tag or interface within a long communication range up to 160 ft. Besides, Bluetooth tags can be used continuously for one year without changing the battery \cite{Tile,StickNFind} by adopting the Bluetooth Low Energy (Bluetooth LE) technique, and they only cost several dollars which are often negligible in comparison with the value of lost objects. In the lost-and-found context, a cheap and miniature Bluetooth tag can be attached to every valuable object and contain its owner's identification information. Once finding his object missing, the owner can use his mobile device to search for the corresponding tag. If the tag gets queried, it can report its location or sound an alert to be located. There are growing commercial Bluetooth-based products for locating personal assets, such as Tile \cite{Tile}, BlueBee \cite{BlueBee}, and StickNFind \cite{StickNFind}. These attractive products, however, often require that a lost object be sufficiently close to the searching device. For example, BlueBee tags \cite{BlueBee} and StickNFind tags \cite{StickNFind} support up to 160 ft and 100 ft, respectively. This inherent range limitation makes it infeasible to recover the lost objects far away from their owners.

%For example, Tile tags \cite{Tile}, BlueBee tags \cite{BlueBee}, and StickNFind tags \cite{StickNFind} support up to 150 ft, 160 ft, and 2,500 ft, respectively. This inherent range limitation makes it infeasible to recover the lost objects far away from their owners.}
\begin{comment}
The plummeting costs of RFID tags make them very promising to revolutionize the lost-and-found service. Specifically, a cheap and miniature RFID tag can be attached to every valuate object and contain its owner's identification information. Once finding his object missing, the owner can use his mobile device to search for the corresponding tag. If the tag gets queried, it can report its location or sound an alert to be located. There are growing commercial RFID products for locating personal assets, such as Tile \cite{Tile}, BlueBee \cite{BlueBee}, and StickNFind \cite{StickNFind}. These attractive products, however, unanimously require that a lost object be sufficient close to the searching device. For example, Tile tags \cite{Tile}, BlueBee tags \cite{BlueBee}, and StickNFind tags \cite{StickNFind} support up to 150 ft, 160 ft, and 2,500 ft, respectively. This inherent range limitation makes it infeasible to recover the lost objects far away from their owners.
\end{comment}
A promising solution to overcoming the above range limitation is via mobile crowdsourcing, which refers to the practice of obtaining needed services or data by soliciting contributions from many mobile users. The emergence of mobile crowdsourcing is driven by the skyrocketing growth of mobile devices. For example, the number of mobile-connected devices would exceed the world population in 2013 and hit 10 billion in 2016 \cite{Cisco2012}. Ubiquitous mobile devices can jointly sense and interact with the physical world at an unprecedented scale, thus enabling many otherwise infeasible applications \cite{SecurZha13,JointShi15,SecurZha15}. One can imagine a service provider offering the object-finding service. An object owner submits an object-finding request as a tag query to the service provider, which in turn forwards the query to selected mobile users referred to as \emph{mobile detectors} hereafter. Every detector then locally broadcasts the query. The tag on the lost object responds to any such query, and the corresponding detector finally sends the tag response and his own location via the service provider to the object owner. Every mobile detector can be rewarded at a fixed rate or in commensurate with the object value. Although the object owner may have to pay for the service, he can recover his valuable object with overwhelming probability.

Crowdsourcing the lost-and-found service faces some great challenges. First, the object in search may be of high value so that the mobile detector discovering it may want to keep it instead of reporting its whereabout to the service provider. Thus we need to alleviate the security concerns of the owners about their lost objects. Second, mobile users may be unwilling to disclose their locations which may indicate too much personal information. Therefore, we must protect the location privacy of mobile users to stimulate their participation in the lost-and-found system. Last, both Bluetooth tags and mobile devices are resource-constrained, so the object-finding process should be very efficient in computation and communication, especially for energy-constrained mobile detectors \cite{HeOn10}. Although some companies such as Tile \cite{Tile} and BlueBee \cite{BlueBee} are offering the crowdsourced lost-and-found service, they ensure neither object security nor location privacy of involved mobile detectors.

This paper presents SecureFind, a crowdsourced object-finding system that offers strong object security to the object owner and also strong location privacy to mobile detectors. The essential idea in SecureFind is to let some mobile detectors generate dummy tag responses which are indistinguishable from the real tag response in the eye of the service provider and other mobile detectors. Only the object owner can identify the real tag response, so strong object security can be ensured. In addition, the location of each mobile detector discovering the lost object is kept from the service provider and only disclosed to the object owner under a dynamic pseudonym. So the location privacy of mobile detectors can be well guaranteed.

Our contributions are mainly threefold. First, we are the first to formulate secure and privacy-preserving object finding via mobile crowdsourcing to the best of our knowledge. Second, we propose two solutions to this problem. The basic scheme provides strong object security at the cost of low efficiency. In contrast, the advanced scheme seeks to achieve a middle ground among object security, location privacy, and energy efficiency. Finally, we thoroughly evaluate the performance of our schemes by theoretical analysis and extensive simulations.

%The rest of this paper is organized as follows. Section~\ref{sec:ProblemStatement} outlines the system model, the adversary model, our design objectives, and a Framed Slotted ALOHA protocol underlying our design. Section~\ref{sec:RelatedWork} surveys the most related work. Section~\ref{sec:BasicScheme} illustrates a basic scheme. Section~\ref{sec:AdvancedScheme} presents an advanced scheme. Section~\ref{sec:Evaluation} evaluates the two schemes using simulations. Section~\ref{sec:Conclusion} concludes this paper.

%%%%%%%%%%%%%%%%%%%%%%%%%%%%%%%%%%%%%%%%%%%%%%%%%%%%%%%%%%%%%%%
\section{Preliminaries}\label{sec:ProblemStatement}
%%%%%%%%%%%%%%%%%%%%%%%%%%%%%%%%%%%%%%%%%%%%%%%%%%%%%%%%%%%%%%%
%In this section, we introduce the system model, the adversary model, the design objectives, and a Framed Slotted ALOHA protocol underlying our scheme design.
\subsection{System Model}\label{subsec:SystemModel}

We assume a SecureFind service provider offering the object-finding service via mobile crowdsourcing. The service provider fulfils every object-finding request through a number of mobile users referred to as \emph{mobile detector}s hereafter. Every detector has a mobile device such as a smartphone or tablet to communicate with the service provider and also nearby Bluetooth tags. Almost all mobile devices are having the Bluetooth functionality, and it has been shown in \cite{YangEsm10} that Bluetooth devices can communicate with each other without explicitly establishing a connection. In addition, nearby mobile detectors can communicate via WiFi-direct, LTE-A, or other available Device-to-Device (D2D) technologies which are widely used in many other applications \cite{YangEsm10,SunPri13,SunSyn14}.
\begin{comment}
An object owner refers to a person who lost a valuable object. We assume that the lost object has an embedded Bluetooth tag hard to find or remove without breaking the object. So we use ``lost tag'' and ``lost object'' interchangeably henceforth. A Bluetooth tag \cite{Tile,StickNFind} is a device which has an on-board battery and can communicate with nearby mobile devices.
\end{comment}

An object owner refers to a person who lost a valuable object. We assume that the lost object is attached with a Bluetooth tag hard to remove without breaking the object and use ``lost tag'' and ``lost object'' interchangeably henceforth. A Bluetooth tag is a small piece of device with an on-board battery, which can perform simple computation and communicate with nearby mobile devices via Bluetooth. Several off-the-shelf Bluetooth tags are currently commercially available for personal asset tracking, such as Tile \cite{Tile}, StickNFind \cite{StickNFind}, and BlueBee \cite{BlueBee} tags. The cost of a Bluetooth tag is currently around a few dollars \cite{Tile} and is plummeting due to rapid technological advance and growing market demand. It is thus reasonable to assume that every high-value object will be attached with a Bluetooth tag to enable object finding in the near future. Moreover, we assume that every tag $i$ has a unique ID $ID_i$ known only to its owner.
%There are many off-the-shelf Bluetooth tags for personal asset tracking,  Finally, we assume that every tag $i$ has a unique ID $ID_i$ known only to its owner. \textcolor{red}{Bluetooth connection before communication.}

\begin{comment}
There are three possible approaches for the server to communicate with the Bluetooth tags. The most intuitive way is to only let the tag attached to lost object respond to the objecting finding request. Although efficient, it guarantees no object security cause it nearby mobile detector knows the existence of the lost object. The second way is to let multiple tags send message to server via mobile detector. However, it suffers from efficiency and scalability problems. A much better alternative is to adopt Framed Slotted Aloha protocol which can greatly improve the energy efficiency since each Bluetooth tag only needs to have one-bit response to each query without incurring too much computation and communication overhead to energy-constraint Bluetooth tags. The Bluetooth tag is much more powerful and can easily support this kind of usage. Besides, the cost of Bluetooth tags is plummeting due to rapid technological advance and growing market demand. It is also reasonable to assume Bluetooth tags for high-value objects. There are many off-the-shelf Bluetooth tags for personal asset tracking, such as Tile \cite{Tile}, StickNFind \cite{StickNFind}, and BlueBee \cite{BlueBee} tags. Finally, we assume that every tag $i$ has a unique ID $ID_i$ known only to its owner.
\end{comment}

The object-finding service in SecureFind works as follows. Assume that the object owner knows that his lost object is likely in a possibly large \emph{target area}, e.g., lower Manhattan. He submits to the service provider an object-finding request containing some information about the lost tag and also the target area. The service provider then forwards the object-finding request to all mobile detectors in the target area, each of which in turn locally broadcasts the request. The lost tag responds to any object-finding request intended for it. Every detector hearing a tag response forwards it and his own location via the server to the object owner. Based on the tag responses, the object owner can derive an approximate location (area) of his lost object, e.g., by multilateral triangulation. Finally, the object owner can go to the derived location and send a tag query in person, in which case the lost tag can respond with its GPS location like a SticknFind tag \cite{StickNFind} or sound an alert like a Tile \cite{Tile} or BlueBee \cite{BlueBee} tag. During this process, the object owner may initiate multiple requests to keep track of the dynamic locations of his lost tag (object) which may be carried and in motion. All the system operations are automatically executed without user involvement through an SecureFind app installed on each mobile device.

Sound incentives must be provided to all the involved parties to materialize SecureFind. The service provider can either charge the object owner at a rate commensurate with the object value, and it may also provide free services and profit by web advertisement when its service goes very popular. Every mobile detector can be rewarded either at a fixed rate or in accordance with the object value. Such rewarding mechanisms as perks or badges have been proved to be very successful in soliciting mobile users for crowdsourcing applications like Foursquare. The object owner may need to pay for the service, but he will be able to quickly recover his lost object of high value.
Here we assume the existence of such incentive mechanisms and refer readers to existing rich literature such as \cite{YangCro12,ZhaoHow14} for incentive design for mobile crowdsourcing.

\subsection{Adversary Model}\label{subsec:AdversaryModel}

We assume that the service provider is honest-but-curious (HBC) \cite{Fountations}, which is a widely adopted assumption for rational service providers. In particular, the service provider is trusted to faithfully follow the protocol execution, but it may have interest in the location of the lost object and also the locations of mobile detectors. In addition, the service provider does not collude with any object owner or mobile detector.

Mobile detectors are curious and also location-sensitive. By curious, we mean that mobile detectors try to locate the lost object and take it away prior to the object owner's arrival. To do so, mobile detectors may attempt to infer whether the lost object is in their vicinity from the information they receive during protocol execution. By location-sensitive, we mean that mobile detectors do not want any party (including the server) to know their accurate locations or equivalently linking their accurate locations to their real IDs.

How to deal with other possible attacks on SecureFind is beyond the scope of this paper. For example, an attacker may jam all radio transmissions, replay intercepted messages, and/or inject bogus messages. Such denial-of-service attacks can target any wireless/mobile system like SecureFind and can be mitigated by existing anti-jamming communication techniques and message authentication.

\subsection{Design Objectives}\label{subsec:DesignObjectives}

We have the following major design objectives.
\begin{itemize}
\item \emph{Correctness}: The object owner should be able to obtain an approximate location of the lost object.
\item \emph{Object security}: The location of the lost object should be known to the object owner only.
\item \emph{Location privacy}: The mapping between the real ID and location of every mobile detector should be kept from any other party.
\item \emph{Efficiency}: The object-finding process should incur low communication and computation overhead.
\end{itemize}

\subsection{Framed Slotted ALOHA Protocol}\label{subsubsec:SlottedALOHA}

Our schemes depend on Framed Slotted ALOHA, which is a popular anti-collision MAC protocol adopted by many RFID systems \cite{TanHow08,TanEff10,LiIde10,ShahCar11,ZhangAss13}. Since Bluetooth tag is much more powerful than RFID tag, it is reasonable to assume that Bluetooth tag can support Framed Slotted ALOHA with minimal modification. In SecureFind, Framed Slotted ALOHA is executed between one mobile detector and a number of nearby Bluetooth tags and works as follows. First, the mobile detector broadcasts a request with two parameters $\langle r,f\rangle$, where $r$ is a random number, and $f$ is the number of time slots in one frame where the $f$ slots are numbered from $0$ to $f-1$. Upon receiving the request $\langle r,f\rangle$, each tag $i$ responds in slot $h(ID_i|| r)\mod f$, where $ID_i$ denotes the unique ID of tag $i$, and $h(\cdot)$ denotes a publicly known hash function. Each of the $f$ time slots can then be an \emph{empty} slot without any tag response, a \emph{singleton} slot with a single tag response, or a \emph{collision} slot with more than one tag responses.

%\begin{table}[t]%\footnotesize
%\renewcommand{\arraystretch}{1.1}
%\caption{Key System Parameters} \centering
%\begin{tabular}{|c|l|}
%\hline \textbf{Para.} &\textbf{Meaning}\\
%\hline
%$f$&  frame length in Frame Slotted ALOHA\\
%$k$&  \# of hash functions\\
%$\omega$& \# of polled positions\\
%$q$& probability that a mobile detector \\
%$$&acts as a dummy tag\\
%$\mathsf{d}_{x,j}$&$j$th polling position in the $x$th polling round\\
%$t$& \# of polling rounds\\
%
%\hline
%\end{tabular}\label{tab:notation}
%\vspace{-.2in}
%\end{table}

%
%\subsubsection{Bloom Filter}\label{subsubsec:Bloomfilter}
%
%A Bloom filter is a space-efficient probabilistic data structure for set-membership testing and many other applications \cite{BrodeNet02,SunPri13}. Assume that we want to use a $u$-bit Bloom filter for a data set $\{s_{i}\}_{i=1}^{d}$, which has every bit initialized to bit-0. Let $\{h_{a}(\cdot)\}_{a=1}^{k}$ denote $k$ different hash functions, each with output in $[1,u]$. Every element $s_{i}$ is added into the Bloom filter by setting all bits at positions $\{h_{a}(s_{i})\}_{a=1}^{k}$ to bit-1. To check the membership of an arbitrary element $e$ in the given data set, we can simply verify whether all the bits at positions $\{h_{a}(e)\}_{a=1}^{k}$ have been set. If not, $e$ is certainly not in the data set; otherwise, it is in the data set with some probability jointly determined by $d,u,$ and $k$.

%%%%%%%%%%%%%%%%%%%%%%%%%%%%%%%%%%%%%%%%%%%%%%%%%%%%%%%%%%%%%%%
\section{Related Work}\label{sec:RelatedWork}
%%%%%%%%%%%%%%%%%%%%%%%%%%%%%%%%%%%%%%%%%%%%%%%%%%%%%%%%%%%%%%%
%To the best of our knowledge, there is no work closely related work to this paper. Therefore, in this section, we briefly discuss some work loosely related to PriFind.

%
%TPC Alex Liu's publication \cite{ShahzEve12,ShahzPro13}
%
%Li Mo \cite{ZhengPET11,ZhengFas11,ZhengZOE13}
%
%Tan Qiu \cite{HanCou10}
%
%Shaojie Tang and Xiang-Yang Li \cite{TangRAS09}
%
%Jiming Chen \cite{SunSen12}
%
%Mario Gerla \cite{LeeRFG09}
%
%Sheng Bo \cite{XieEff10,HanCou10}

%To the best of our knowledge, SecureFind is the first work on secure and privacy-preserving crowdsourced search for lost objects. The following work in RFID systems is loosely related to SecureFind.

Several schemes have been proposed for tracking and locating lost objects. AutoWitness \cite{GuhaAut10} is a personal asset tracking system that uses an embedded tag with inertial sensor to estimate asset's position change and proactively transmit trajectory data to an external server via cellular link to facilitate asset retrieval. In contrast, SecureFind depends on low-cost Bluetooth tags without any inertial sensor or cellular communication capabilities, thus more suitable for wide adoption. Moreover, Sherlock \cite{NemmaShe08} is a system designed to localize objects with embedded RFID tags in some closed space, which cannot be applied to find lost object  in outdoor and is thus orthogonal to SecureFind.

Recent years have witnessed significant research on missing-tag detection \cite{HushAna98,LawEff00,ZhouEva04,MyungAda06,TanHow08,TanEff10,LuoEff11,LuoMis14,LuoPro12} and identification \cite{LiIde10,ZhangFas11,ZhengPMT13} in RFID systems. This line of work aims to quickly detect whether or which tags are missing in a large RFID system, while SecureFind targets a totally different problem. In particular, a lost tag in SecureFind is a tag lost by its owner but still in the SecureFind service provider's service region, and SecureFind aims to determine which mobile detector has the lost tag in his coverage in order to locate and retrieve the lost object without revealing such information to either the mobile detector or service provider. In contrast, a missing tag in \cite{HushAna98,LawEff00,ZhouEva04,MyungAda06,TanHow08,TanEff10,LuoEff11,LuoMis14,LuoPro12,LiIde10,ZhangFas11,ZhengPMT13} means a tag taken away from the monitored area, and the goal there is to determine if any tag is missing. Therefore, existing missing-tag detection schemes are inapplicable to our problem.

Also related is the line of work on privacy-preserving tag identification and authentication in RFID systems, e.g., \cite{YangIde10,LiPri12,LuAct09,YaoPer09,AlomaSca10,TancSec08}. These schemes allow efficient identification and authentication of an RFID tag without disclosing any information that can be used to uniquely identify the tag. All the RFID tags belong to the same administrator, and there is no attempt to hide the locations of the RFID tags from the administrator. In contrast, each Bluetooth tag in SecureFind belongs to the corresponding object owner, and its location should be protected from the service provider as well. Therefore, SecureFind differs significantly from these schemes in its aim and scope.

\section{A Basic Scheme}\label{sec:BasicScheme}
%%%%%%%%%%%%%%%%%%%%%%%%%%%%%%%%%%%%%%%%%%%%%%%%%%%%%%%%%%%%%%%
In this section, we present a basic scheme for secure and privacy-preserving object finding. The essential idea is to let some mobile detectors in the target area act as \emph{dummy tags} to send dummy tag responses for concealing the real tag response. Since the mobile detectors near the lost object cannot differentiate between real and dummy tag responses, the security of the lost object can be well protected. The major design challenge here is how to let the object owner discover the mobile detectors close to his lost object without drawing the attention of these mobile detectors or the service provider.

We propose an iterative multi-round protocol as a solution. In each round, each mobile detector executes the Framed Slotted ALOHA protocol in Section~\ref{subsubsec:SlottedALOHA} and forwards the execution result to the object owner via the service provider. The object owner then excludes some mobile detectors who are unlikely near his lost object according to their execution results. The protocol completes when no more mobile detectors can be excluded. Finally, the object owner retrieves the locations of the remaining mobile detectors from the server provider using some specific cryptographic technique and then infer the location of his lost object. Our scheme ensures that neither the service provider nor the remaining mobile detectors can learn the location of the lost object.

\subsection{Scheme Description}

First, the object owner submits an object-finding request $\langle H(\widehat{ID}||r),r,\textsf{PK}\rangle$ and the target area to the service provider, where $\widehat{ID}$ denotes the ID of the lost tag, $r$ is a random seed, $H(\cdot)$ denotes a publicly known cryptographic hash function, and $\textsf{PK}$ is the object owner's public key. We can also replace $\textsf{PK}$ with a public-key certificate to prevent the service provider from changing PK to its own choice.  We assume that the service provider knows the physical zone each mobile detector resides but not his accurate location. Upon receiving the request, the service provider forwards the request to all the mobile detectors in the target area, each of which then locally broadcasts a tag query $\langle H(\widehat{ID}||r),r\rangle$. Here we assume a suitable MAC protocol to resolve potential collisions among mobile detectors; e.g., each mobile detector can wait for some random time before sending the tag query. Every tag seeing such a tag query can check whether it is the intended tag by comparing the hash over its ID and $r$ with the received one, and only the lost tag gets prepared to respond. In addition, each mobile detector returns his location encrypted with $\textsf{PK}$ to the service provider so that the service provider cannot figure out his accurate location. The service provider temporarily buffers these encrypted locations.

The object owner then initiates a polling phase consisting of multiple rounds. Consider round $x\geq 1$ as an example. The object owner sends a polling request $\langle r_x,f\rangle$ via the service provider to each mobile detector, where $f$ denotes the frame length as a fixed system parameter, and $r_x$ is a fresh random seed. Every detector $i$ then locally broadcasts $\langle r_x,f\rangle$. Every other detector hearing the polling request from detector $i$ chooses himself as a dummy tag with probability $q$, which is a tunable system parameter given by the service provider. Each dummy tag $j$ also generates a random pseudonym $ID_j$. Let $\mathcal{T}_{x,i}$ denote a set of tags comprising all the dummy tags near detector $i$ and also the lost tag if it hears the polling request from detector $i$ as well. Let $h_1(\cdot),\dots, h_k(\cdot)$ be $k$ publicly known hash functions, where $k$ is a system parameter. Every tag $j\in \mathcal{T}_{x,i}$ computes $k$ slots to reply, where the $\alpha$th slot is computed as $\mathsf{s}^\alpha_{j,x}=h_\alpha(ID_j||r_x)\mod f$ for all $\alpha\in[1,k]$. During the execution of Framed Slotted ALOHA, every tag $j$ sends a one-bit short response in each of its $k$ computed slots. In the end of round $x$, detector $i$ obtains a bit vector $\mathsf{V}_{i,x}=\langle v_{i,x}[0],\dots,v_{i,x}[f-1]\rangle$, where $v_{i,x}[y]=0$ if slot $y$ is an empty slot and $v_{i,x}[y]=1$ otherwise. Note that here we do not differentiate between singleton and collision slots, which would require each tag to reply a long multi-bit response and thus incur higher communication overhead. Then detector $i$ sends its bit vector $\mathsf{V}_{i,x}$ to the object owner via the server.

Assuming that there are totally $C$ mobile detectors in the target area, the object owner receives $C$ bit vectors $\{\mathsf{V}_{i,x}\}^C_{i=1}$ in round $x$. He then checks if any mobile detector can be excluded, which is certainly not in the transmission range of his lost tag. To do so, the object owner maintains a candidate detector set. Let $\mathcal{C}_x$ be the candidate detector set at the beginning of round $x$, where $\mathcal{C}_1=\{1,\dots,C\}$.
%Here we assume that the service provider has allocated each detector a unique pseudonym for this service request and notified the object owner about it as well. At the end of round $x$, the object owner updates $\mathcal{C}_x$ as follows.
For each detector $i\in\mathcal{C}_x$, the object owners checks if at least one of the bit positions (or slots) $\{h_\alpha(\widehat{ID}||r_x)\mod f\}^k_{\alpha=1}$ in $\mathsf{V}_{i,x}$ is zero (or empty), where $\widehat{ID}$ is the ID of his lost tag. If so, the lost tag is certainly not around detector $i$, and no dummy tag replied in that slot either. So detector $i$ can be safely removed from $\mathcal{C}_x$. The object owner terminates the polling phase if the number of candidate detectors drops to one or remains unchanged after $\tau\geq 2$ polling rounds, where $\tau$ is a system parameter. The latter case occurs when the lost tag lies in the coverage of multiple detectors. Also note that the candidate detector set remains confidential to the object owner, and all the $C$ mobile detectors need to broadcast the polling request and process the responses in each round of the polling phase even if some of them may have been confidentially excluded by the object owner.

Once the polling phase is over, the object owner retrieves the encrypted locations of the remaining candidate detectors from the service provider. Finally, he can derive an approximate range for his lost object based on the decrypted detector locations. We can see that the service provider will know which mobile detectors are not excluded. Since the service provider knows the physical zone each mobile detector resides (instead of his real location), it can deduce that the lost object is in one of the physical zones of the remaining detectors. There are two ways to alleviate this security concern. First, the object owner can request the encrypted locations of $c\geq 1$ detectors that include both the remaining detectors and some excluded detectors to confuse the service provider. Second, the object owner can execute an efficient Private-Information-Retrieval protocol \cite{AsharMor13} to retrieve the encrypted locations of the remaining candidate detectors without revealing whose locations are retrieved.

\subsection{Performance Analysis}\label{sec:analysis-basic}
Now we analyze the performance of the basic scheme.
%The enhanced scheme offers the same level of strong location privacy to mobile detectors as the basic scheme.
\begin{comment}
\textcolor{red}{In addition, we will resort to simulations to evaluate the object security offered by the enhanced scheme, which is quite involved to theoretically analyze.}
\end{comment}

\noindent \underline{\textbf{Correctness.}}\hspace{.05in}  The basic scheme can guarantee that the object owner obtains an approximate location for his lost object as long as it is within the transmission range of at least one mobile detector. Assume that there are totally $N$ mobile users in a region of area $S$. Also suppose that the number of mobile detectors in any subregion of area $s$, denoted by $X(s)$, follows a homogeneous spatial Poisson process with intensity $N/S$: $\mathsf{Pr}(X(s)=k)=\frac{(Ns/S)^ke^{-Ns/S}}{k!}$. Let $R$ denote the transmission range of the lost tag and also mobile detectors. It is easy to see that the basic scheme is correct with probability $1-\mathsf{Pr}(X(\pi R^2)=0)=1-e^{-\pi N R^2/S}$.

In addition, the basic scheme may incur false positives, which occur when the lost object is not close to any mobile detector (i.e., the given target area is wrong), but some dummy tags happen to respond just like the lost tag in each round of the polling phase. The object owner thus will be misled to wrong locations. We can estimate the false-positive probability as follows. Consider any of the $C$ detectors in the target area, say detector $i$, which has on average $c = \lfloor\pi N R^2/S\rfloor$ other mobile detectors in his transmission range and does not have the lost tag $\widehat{ID}$ there. Since each mobile detector acts as a dummy tag with probability $q$, there are totally $cq$ dummy tags in detector $i$'s coverage. Recall that the lost tag needs to respond in slots $\{\mathsf{s}^\alpha_{j,x}=h_\alpha(ID_j||r_x)\mod f\}_{\alpha=1}^k$ in round $x$ if hearing a polling request. Assume that the output of every hash function is uniformly distributed in $[0,f-1]$. Then the average number of distinct slots the lost tag needs to respond is given by
\begin{equation}\label{eq:mu}
\mu=\sum_{l=1}^k {l\times \frac{{f\choose l}}{f^k}}\;.
\end{equation}
As said, each dummy tag also responds in up to $k$ slots uniformly distributed in $[1,f]$.
The probability that no dummy tag responds in a particular slot of the lost tag is given by $(1-1/f)^{kqc}$. For detector $i$ to stay in the object owner's candidate detector set in round $x$, at least one dummy tag needs to respond in each of the $\mu$ distinct slots, which occurs with probability $p_\textrm{one}=\big(1-(1-1/f)^{kqc}\big)^\mu$. Assume that the polling phase terminates in $t$ rounds. For the false positive to occur, at least one detector needs to survive all the $t$ rounds, which occurs with probability $1-(1-p^t_\textrm{one})^C$.

\vspace{.1in}\noindent \underline{\textbf{Object Security.}}\hspace{.05in}  The basic scheme offers strong object security. In particular, the information the service provider can obtain during object finding includes the initial object-finding request $\langle H(\widehat{ID}||r),r,\textsf{PK}\rangle$, the polling results in each round, and from which candidate detectors the object owner requested the location. Since the service provider knows neither ID of the lost tag nor the random pseudonym of each dummy tag, he cannot directly infer which detectors have the lost tag in their coverage from the polling results besides knowing that one of the detectors for which the object owner requested the locations does.

Can the service provider do better? To make quantitative analysis possible, we assume that the average number of tags in each detector's communication range are the same, e.g., $cq$. Under this assumption, the detector with the lost tag in its coverage may observe slightly more non-empty slots than those without during the polling phase. In particular, each detector covering the lost tag, called a real detector hereafter, observes a non-empty slot in each slot with probability $p_1=1-(1-1/f)^{(cq+1)k}$, whereas each detector not covering the lost tag, called a fake detector hereafter, does so with probability $p'_1=1-(1-1/f)^{cqk}$. Although this is only a rough estimate because the number of dummy tags around each mobile detector are most likely different, the service provider may still try to gain some information from the polling results by ranking all the detectors according to the numbers of bit ones in their reported vectors. More specifically, the higher the rank of a detector (i.e., the more bit ones in  reported vectors), the more likely the detector is a real one, and vice versa.

Now we analyze the probability distribution of the real detector's rank. Consider a real detector $i$ and a fake detector $j$ in round $x$ as an example. Denote by $b_i$ and $b_j$ the numbers of bit-one positions in their reported vectors $\mathsf{V}_{i,x}$ and $\mathsf{V}_{j,x}$, respectively. Let $u=\min(f,(cq+1)k)$ and $u'=\min(f,cqk)$. The probability that detector $i$ has  more bit-one positions than detector $j$ is given by
\begin{equation}\label{eq:prob}
\begin{split}
p_m&=\mathsf{Pr}(b_i\geq b_j)\\
 &=\sum^{u'}_{z=0}\mathsf{Pr}(b_i\geq z)\cdot \mathsf{Pr}(b_j=z)\\
&=\sum^{u'}_{z=1}\sum^u_{z'=z}\mathsf{Pr}(b_i=z')\cdot \mathsf{Pr}(b_j=z)\\
&=\sum^{u'}_{z=1}\sum^u_{z'=z} {u\choose z'}p_1^{z'}(1-p_1)^{u-z'}{u'\choose z}{p'}_1^{z}(1-{p'}_1)^{u'-z}\;.\\
%&=\sum^{u'}_{z=1}\sum^u_{z'=z} {u\choose z'}{u'\choose z}p_1^{z'+z}(1-p_1)^{u+u'-z-z'}\;.
\end{split}
\end{equation}

For simplicity,  assume that there is only one real detector. The p.d.f. of real detector's rank is then given by
\begin{equation}\label{equ:rank}
\mathsf{Pr}(\textrm{rank}=r) = {{C-1}\choose {r-1}}p_m^{r-1}(1-p_m)^{C-r}\;.
\end{equation}
We can see from Eqs.~(\ref{eq:prob}) and (\ref{equ:rank}) that if the number of dummy tags (i.e., $cq$) is large, $p_1$ is very close to $p'_1$. This means that the real detector will be ranked in the middle of all the detectors with high probability, and the object security can thus be guaranteed.

%Second, due to the use of Private Information Retrieval \cite{AsharMor13}, the service provider cannot figure out which mobile detectors' encrypted locations the object owner finally retrieves. Therefore, although the service provider knows the physical zone of each mobile detector, it cannot infer the physical zone the lost object might be in. In other words, the service provider only knows the big region the lost object may be in after the protocol execution, which it has been told by the object owner before the protocol execution.

In addition, neither true or fake mobile detectors can distinguish the responses from the lost tag and from dummy tags and thus cannot determine whether the lost tag is in its vicinity.

\begin{comment}
\textcolor{red}{Let us consider an extreme situation. Assume that detector $i$ has the lost object in its coverage. Recall that the lost tag responds in up to $k$ slots in every round. If detector $i$ finds no more than $k$ busy slots in each of the $t$ rounds, there are three possible cases: (1) $m\geq 2$ tags (including dummy tags and/or the lost tag) responded and chose at most $k$ distinct slots in total in every round; (2) only one dummy tag responded in every round; and (3) only the lost tag responds in every round. Since each of the $c$ neighboring detectors of detector $i$ acts as a dummy tag with probability $q$, we can easily estimate that the first case occurs with probability $\big(\sum_{l=1}^k \frac{{f\choose l}}{f^{mk}}\big)^t$, the second case happens with probability $cq(1-q)^{c-1}$, and the third case happens with probability $(1-q)^{c}$. Detector $i$ can then compare the three probability values and decides whether to search for the lost object.}
\end{comment}

\vspace{.1in}\noindent \underline{\textbf{Location Privacy.}}\hspace{.05in}  The basic scheme offers strong location privacy to mobile detectors. Specifically, each mobile detector can report a physical zone encompassing his location instead of his real location to the service provider to participate in SecureFind. Therefore, the service provider cannot get the accurate location of any detector. Even if the location of every responding detector is disclosed to the object owner, we can hide the real ID of the detector from the object owner by letting the service provider replace the real ID with a dynamic pseudonym. Since the objector owner does not collude with the service provider per our adversary model, the location privacy of every mobile detector is well preserved.

\vspace{.1in}\noindent \underline{\textbf{Efficiency.}}\hspace{.05in} To analyze the communication overhead of the basic scheme, we first derive the expected number $t$ of polling rounds. For any mobile detector not covering the lost tag, the object owner excludes it from the candidate detector set with probability
\[
p_e=1-p_\textrm{one}=1-\big(1-(1-1/f)^{kqc}\big)^\mu\;,
\]
where $\mu$ is given in Eq.~(\ref{eq:mu}). So the object owner can exclude $p_{\textrm{e}}$ fraction of the remaining candidate detectors after each polling round. Assume that the number of candidate detectors drops to one after $t$ rounds. Then we have $Cp_{\textrm{e}}^t= 1$ and thus
\begin{equation}\label{eq:rounds}
t=\lfloor\log_{p_{\textrm{e}}}\frac{1}{C}\rfloor\;.
\end{equation}
Each mobile detector sends its encrypted location to the service provider at the beginning, and he also broadcasts a polling request and sends a $f$-bit vector to the service provider in each polling round. In addition, since each tag needs to reply $k$ one-bit responses in each round, the total communication overhead incurred by tag responses is about $cktC$ bits. Moreover, the object owner sends one object-finding request and $t$ polling messages. Finally, the object owner retrieves $\lambda$ encrypted detector locations from the service provider.

As for the computation overhead, each tag (dummy or lost) needs $k$ efficient hash operations in each polling round, leading to $cktC$ hash operations in total. Moreover, each mobile detector performs one public-key encryption, and the object owner needs to carry out one public-key decryption for each non-excluded mobile detector. The most expensive public-key encryptions and decryptions can be done very efficiently on current mobile devices. For example, for the standard Elliptic Curve Integrated Encryption Scheme (ECIES), one point multiplication and two point multiplications are needed for one decryption and one encryption, respectively, and a point multiplication takes less than 7.3 ms on an Android Galaxy Nexus smartphone \cite{FanSec13}.

\section{An Advanced Scheme: Selected Polling}\label{sec:AdvancedScheme}
%%%%%%%%%%%%%%%%%%%%%%%%%%%%%%%%%%%%%%%%%%%%%%%%%%%%%%%%%%%%%%%
The basic scheme provides strong object security. However, in each polling round, each mobile detector needs to send an $f$-bit vector to the service provider which incurs large communication overhead and low efficiency. In this section, we present an advanced scheme to strike a middle ground between object security and system efficiency.

\subsection{Basic Idea}

The advanced scheme stems from an observation about the basic scheme. Specifically, the response from every detector in each polling round is an $f$-bit vector. The object owner excludes some candidate detectors in each round $x$ by checking the bit values at $k$ positions $\{\mathsf{s}^\alpha_{j,x}=h_\alpha(ID_j||r_x)\mod f\}_{\alpha=1}^k$, which we refer to as \emph{real} positions. There are at most $k$ real positions because some modular hash values may be the same. Accordingly, we refer to the rest no less than $f-k$ bit positions as \emph{dummy} positions. The dummy positions can effectively hide the real positions so that the detector with the lost object in its coverage cannot tell. The efficiency can be improved if fewer dummy positions are used in each polling round, and the accompanying cost is that real positions will have a higher chance of exposure.

The advanced scheme implements the above thinking by letting the object owner selectively poll fewer than $f$ bit positions in each round, among which the fraction of real positions is adjusted based on the results in previous polling rounds. Intuitively, the more real positions polled in each round, the fewer polling rounds needed to locate the lost tag, the lower the communication and computation overhead, the higher chance of exposing the lost tag, and vice versa. The challenge is how to characterize the exposure of the lost tag and then properly adjust the fraction of real positions.

What is the impact of polling fewer dummy positions on object security? Consider an arbitrary mobile detector, say $i$. If detector $i$ has the lost tag in his coverage, he is more likely to observe more non-empty slots than other detectors not covering the lost tag. More specifically, assume that the object owner queries $\omega$ out of $f$ bit positions, which consists of $\gamma\geq 1$ real positions and $\omega-\gamma$ dummy positions. Recall that each detector on average has $c=\lfloor \pi R^2 N/S\rfloor$ other detectors in his coverage, each acting as a dummy tag with probability $q$. If detector $i$ covers the lost tag, the probability that a randomly queried bit position having a one (or equivalently the corresponding slot is busy) can be estimated as
\begin{equation}\label{eq:H1}
\begin{split}
p_1&=(1-(1-1/f)^{cqk})\frac{\omega-\gamma}{\omega}+\frac{\gamma}{\omega}\\
&=1-(1-1/f)^{cqk}+(1-1/f)^{cqk}\frac{\gamma}{\omega}\;.
\end{split}
\end{equation}
If the lost tag is outside detector $i$'s coverage, the above probability is $p'_1=1-(1-1/f)^{cqk}$. It is easy to see that $p'_1<p_1$ for $\gamma\geq 1$.  As we normally have $\gamma/\omega > k/f$, the gap between $p_1$ and $p_1'$ becomes more noticeable in the advanced scheme, leading to lower object security. In addition, the larger $\gamma$, the more quickly the object owner ruling out the candidate detectors not covering the lost object, the fewer polling rounds needed, the larger the probability gap, the lower object security, and verse versa.

To strike a balance between object security and system efficiency, we let the object owner maximize the number of real positions in each polling round as long as the polling result (i.e., the $\omega$-bit vector) observed by the detector covering the lost object is \emph{statistically indistinguishable} from the one observed by a detector not covering the lost tag.
More specifically, let the null hypothesis be that the $\omega$-bit vector obtained by a detector is generated from the binomial distribution $B(\omega,p'_1)$, i.e., the theoretical distribution. We can then test the hypothesis using Pearson's chi-squared test \cite{ChernUse54} with the test statistics given by
\begin{equation}
\chi^2 = \frac{(p_{\textrm{ob}}-p'_1)^2}{p'_1} + \frac{((1-p_{\textrm{ob}})-(1-p'_1))^2}{(1-p'_1)}\;,
\end{equation}
where $p_{\textrm{ob}}$ is the observed frequency of bit ones, and $p'_1=1-(1-1/f)^{cqk}$ is the theoretical frequency. Finally, we can compute a $p$-value from $\chi^2$ using the chi-squared distribution for one degree of freedom, which gives us the probability of observing such difference if the $\omega$-bit vector is generated from $B(\omega,p'_1)$.

%\textcolor{red}{needs to use a different notation for $p$, the probability of each detector acting as a dummy tag to avoid confusion with p-value.}

\subsection{Scheme Description}

The pre-polling phase of the advanced scheme is exactly the same as that of the basic scheme, so we do not repeat it here for lack of space.

As in the basic scheme, the polling phase in the advanced scheme also consists of multiple rounds. Consider round $x\geq 1$ as an example. The object owner sends a polling request $\langle r_x, f, \mathsf{d}_{x,0},\dots,\mathsf{d}_{x,\omega-1}\rangle$ via the service provider to each mobile detector, where $f$ denotes the frame length as a fixed system parameter, $r_x$ is a fresh random seed, and $0\leq \mathsf{d}_{x,0}<\mathsf{d}_{x,1}<\dots<\mathsf{d}_{x,\omega-1}\leq f-1$ are the $\omega$ bit positions that the object owner intends to poll in round $x$. These $\omega$ bit positions include $\gamma_x$ real and $\omega-\gamma_x$ dummy positions, and how to choose them will be discussed shortly. Every detector $i$ then locally broadcasts $\langle r_x, f, \mathsf{d}_{x,0},\dots,\mathsf{d}_{x,\omega-1}\rangle$. Every other detector hearing the polling request from detector $i$ chooses himself as a dummy tag with probability $q$ which is a system parameter. Let $\mathcal{T}_{x,i}$ denote the set of tags comprising all the dummy tags near detector $i$ and also the lost tag if it is covered by detector $i$. The Framed Slotted ALOHA protocol is still used to collect tag responses. Every tag $j\in \mathcal{T}_{x,i}$ computes $k$ candidate slots to reply, where the $\alpha$th slot is computed as $\mathsf{s}^\alpha_{j,x}=h_\alpha(ID_j||r_x)\mod f$. Then for each $\mathsf{d}_{x,y},y\in[0,\omega-1]$, tag $j$ checks if $\mathsf{d}_{x,y}=\mathsf{s}^\alpha_{j,x}$ for some $\alpha$. If so, tag $j$ knows that it should reply a one-bit response in slot $y$ and keeps silent otherwise. In the end of round $x$, detector $i$ obtains a $\omega$-bit vector $\mathsf{W}_{i,x}=\langle \mathsf{w}_{i,x}[0],\dots,\mathsf{w}_{i,x}[\omega-1]\rangle$, where $\mathsf{w}_{i,x}[y]=0$ if slot $y$ is an empty slot and $\mathsf{w}_{i,x}[y]=1$ otherwise. Then detector $i$ sends $\mathsf{W}_{i,x}$ to the object owner via the service provider.

Given totally $C$ mobile detectors in the target area, the object owner receives $\{\mathsf{W}_{i,x}\}^C_{i=1}$ in round $x$. As in the basic scheme, he maintains a set of candidate detectors which initially contain all the $C$ detectors. After receiving $\{\mathsf{W}_{i,x}\}^C_{i=1}$, the object owner eliminates all the detectors from the candidate set $\mathcal{C}_x$ with each having at least one zero at the $\gamma_x$ real positions in his polling result. The polling phase stops when the number of candidate detectors drops to one or remains unchanged after $\tau\geq 2$ rounds, where $\tau$ is a system parameter.

After the polling phase, the object owner retrieves the encrypted locations of $\lambda\geq 1$ detectors that include both the remaining detectors and some excluded detectors from the service provider. Finally, he can derive an approximate range for his lost object based on the decrypted detector locations as in the basic scheme.

\subsection{Choosing Polling Positions}

Now we discuss how to choose the $\omega_x$ polling positions $\{\mathsf{d}_{x,j}\}_{j=0}^{\omega-1}$ in each round $x$.

The first step is to determine $\gamma_x$, the number of real positions in round $x$. We propose to derive $\gamma_x$ based on the $C$ polling results received in all previous rounds such that the expected polling results in round $x$ are statistically indistinguishable from the results generated from the theoretical binomial distribution $B(\omega,p'_1)$. In particular, recall that $\mathcal{C}_x$ denote the set of remaining candidate detectors at the beginning of round $x$. Let $b_{i,x-1}$ be the number of bit ones in $\mathsf{W}_{i,x-1}$ for all $i\in\mathcal{C}_x$, where we set $b_{i,0}=\lceil(1-(1-1/f)^{cqk})\omega\rceil$. As discussed, the probability of any bit position in $\mathsf{W}_{i,x}$ being one for any detector $i\in\mathcal{C}_x$ not covering the lost object can be derived as $p_{i,1}=1-(1-1/f)^{cqk}$. Then the object owner tries to find $\gamma_{x,i}$ for each detector $i\in \mathcal{C}_x$, the largest number of real positions can be polled in round $x$, if detector $i$ covers the lost tag. To do so, the object owner initially set $\gamma_{x,i}=0$. According to Eq.~(\ref{eq:H1}), the probability of any bit position in $\mathsf{W}_{i,x}$ being one if detector $i$ covers the lost tag is
\[
\hat{p}_{i,1}=(1-(1-1/f)^{cqk})\cdot\frac{\omega-\gamma_{x,i}}{\omega}+\frac{\gamma_{x,i}}{\omega}\;.
\]
He then computes the expected fraction of bit ones in $\mathsf{W}_{i,x-1}||\mathsf{W}_{i,x}$ as
$p_{\textrm{ob}}=\frac{\hat{p}_{i,1}\omega+b_{i,x-1}}{2\omega}$, the corresponding test statistics $\chi^2$, and finally the $p$-value (denoted by $p_{\textrm{val},i}$). If $p_{\textrm{val},i}>p_{\textrm{thre}}$, where $p_{\textrm{thre}}$ is the threshold chosen by the object owner, he increases $\gamma_{x,i}$ by one and repeats the above process until find the largest possible $\gamma_{x,i}\leq k$. Finally, he chooses $\gamma_x$ as the minimum among $\{\gamma_i|i\in\mathcal{C}_x\}$.
After determining $\gamma_x$, the object owner then constructs $\mathsf{q}_{x,0},\dots,\mathsf{q}_{x,\omega-1}$ by randomly choosing $\gamma_x$ real positions from $\{\mathsf{s}^\alpha_{j,x}\}^k_{\alpha=1}$ and $\omega-\gamma$ dummy positions. The above process is summarized in Algorithm~\ref{algo_disjdecomp}.
%\restylealgo{boxed}\linesnumbered
\begin{algorithm}
\SetKwInOut{Input}{input}
\SetKwInOut{Output}{output}
\caption{Computing $\gamma_x$ for round $x$}
\Input{Bit vectors $\{b_{i,x-1}|i\in\mathcal{C}_x\}$, frame length $f$, $p$-value threshold $p_{\textrm{thre}}$}
\Output{$\gamma_x$: the number of real positions in round $x$}
%\BlankLine
%\emph{special treatment of the first line}\;
$\gamma_x\longleftarrow\min(k,\omega)$\;
\ForEach{$i\in\mathcal{C}_x$}{
$\gamma_{x,i} \longleftarrow 0,p_{\mathrm{val},i}\longleftarrow1$\;
$p_{i,1}\longleftarrow1-(1-1/f)^{cqk}$\;
\While{$p_{\mathrm{val},i}>p_{\mathrm{thre}}$}{
$\hat{p}_{i,1}\longleftarrow(1-(1-1/f)^{cqk})\cdot\frac{\omega-\gamma_i}{\omega}+\frac{\gamma_i}{\omega}$\;
$p_{\mathrm{ob}}\longleftarrow\frac{\hat{p}_{i,1}\omega+b_{i,x-1}}{2\omega}$\;
$\chi^2 = \frac{(p_{\mathrm{ob}}-p_{i,1})^2}{p_{i,1}} + \frac{((1-p_{\mathrm{ob}})-(1-p_{i,1}))^2}{(1-p_{i,1})}$\;
Update $p_{\mathrm{val},i}$ according to $\chi^2$ based on chi-square distribution\;
\If{$p_{\mathrm{val},i}> p_{\mathrm{thre}}$}{
$\gamma_{x,i}\longleftarrow\gamma_{x,i}+1$\;
}
\Else{$\gamma_{x,i}\longleftarrow\gamma_{x,i}-1$\;
}
}
\If{$\gamma_{x,i}<\gamma_x$}{
$\gamma_x\longleftarrow\gamma_{x,i}$\;
}
}
\Return $\gamma_x$\;
\label{algo_disjdecomp}
\end{algorithm}
%\decmargin{1em}
%\end{comment}

\subsection{Performance Analysis}

The advanced scheme is correct with the same overwhelming probability and offers the same level of strong location privacy to mobile detectors as the basic scheme.

%Below we only briefly analyze the object security and the efficiency due to space limitations.

\vspace{.1in}\noindent \underline{\textbf{Object Security.}} Similar to that in the basic scheme, the service provider may rank the detectors based on the number of bit ones in their reported vectors. Since we normally have $\gamma/\omega > k/f$, the gap between $p_1$ and $p_1'$ is more noticeable in the advanced scheme than that in the basic scheme. We thus expect that the advanced scheme offers lower object security than the basic scheme does. Since the number of real positions queried in each polling round is jointly determined by the previous polling results and $p_{\mathrm{thre}}$, we have not been able to derive the rank distribution of the real detector. Instead, we evaluate the object security of the advanced scheme in Section~\ref{sec:Evaluation}.

\vspace{.1in}\noindent \underline{\textbf{Efficiency.}} The communication overhead of the advanced scheme depends on the number of polling rounds. Each mobile detector sends its encrypted location to the service provider at the beginning, and he also broadcasts a polling request and sends a $\omega$-bit vector to the service provider in each polling round. In addition, each tag needs to reply $k\omega/f$ one-bit responses on average in each round, so the total communication overhead incurred by tag response is about $ck\omega tC/f$ bits. Moreover, the object owner sends one object-finding request and $t$ polling messages. Finally, the object owner retrieves  $\lambda$ encrypted detector locations from the service provider.

As for the computation overhead, each tag (dummy or lost) needs $k$ efficient hash operations in each polling round, leading to $cktC$ hash operations in total. Because the number of polled real positions in the advanced scheme is smaller than that in the basic scheme, the number of polling rounds is also larger in the advanced scheme, resulting in more hash operations and thus larger tag computation overhead. Moreover, each mobile detector performs one public-key encryption, and the object owner needs to carry out one public-key decryption for each non-excluded mobile detector. As said, such public-key encryptions and decryptions can be efficiently done on modern mobile devices.

Again, since the number of polling rounds is jointly determined by the previous polling results and $p_{\mathrm{thre}}$, we have not been able to derive a closed-form result for the communication and computation overhead of the advanced scheme, which is evaluated via simulations in Section~\ref{sec:Evaluation}.

\begin{figure*}[t]
\centerline{
\subfigure[Normalized rank]{\label{fig:p-rank}
\includegraphics[width=1.8in]{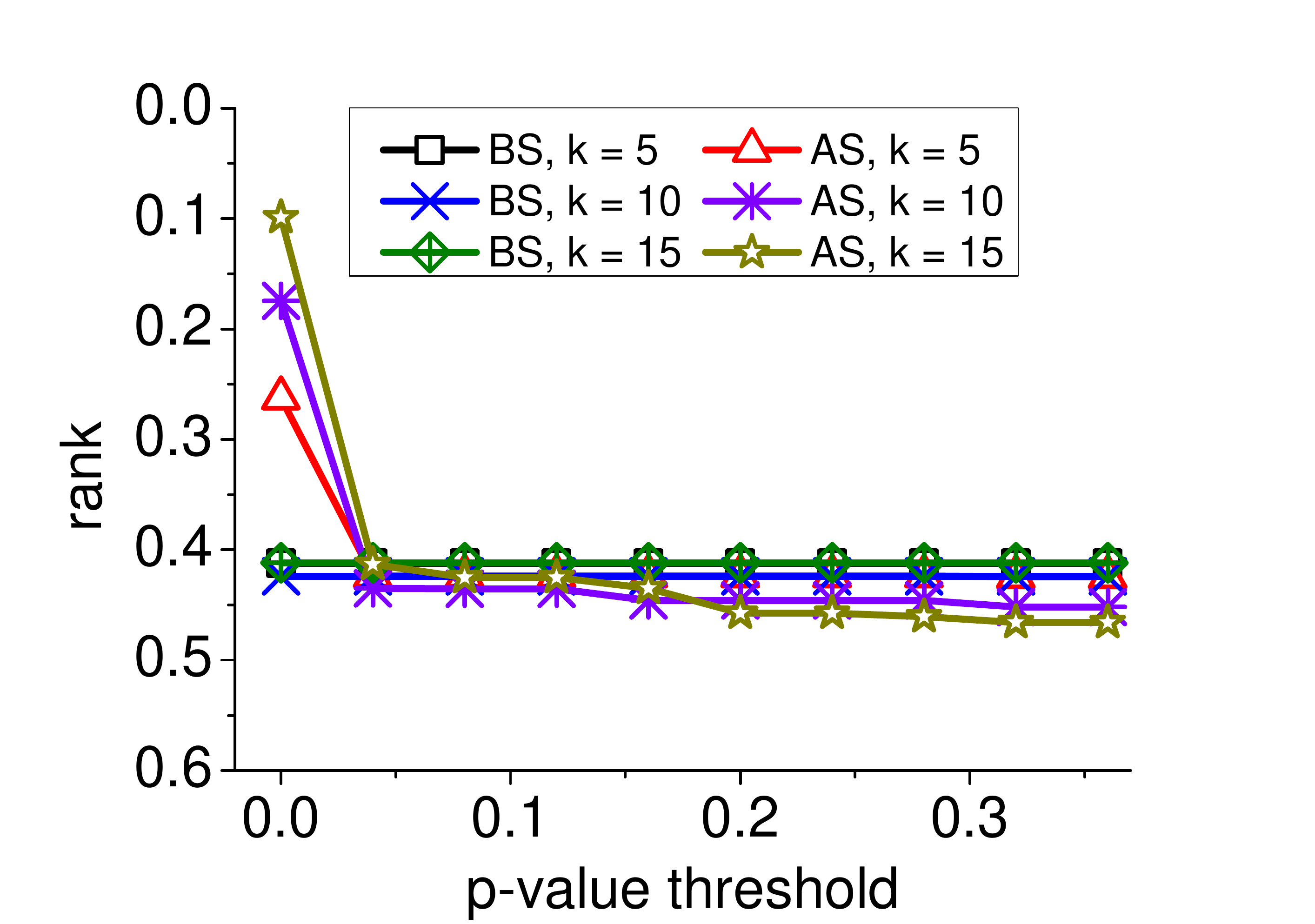}}\hfill
\subfigure[Tag-comm. overhead]{\label{fig:p-tag-comm}
\includegraphics[width=1.8in]{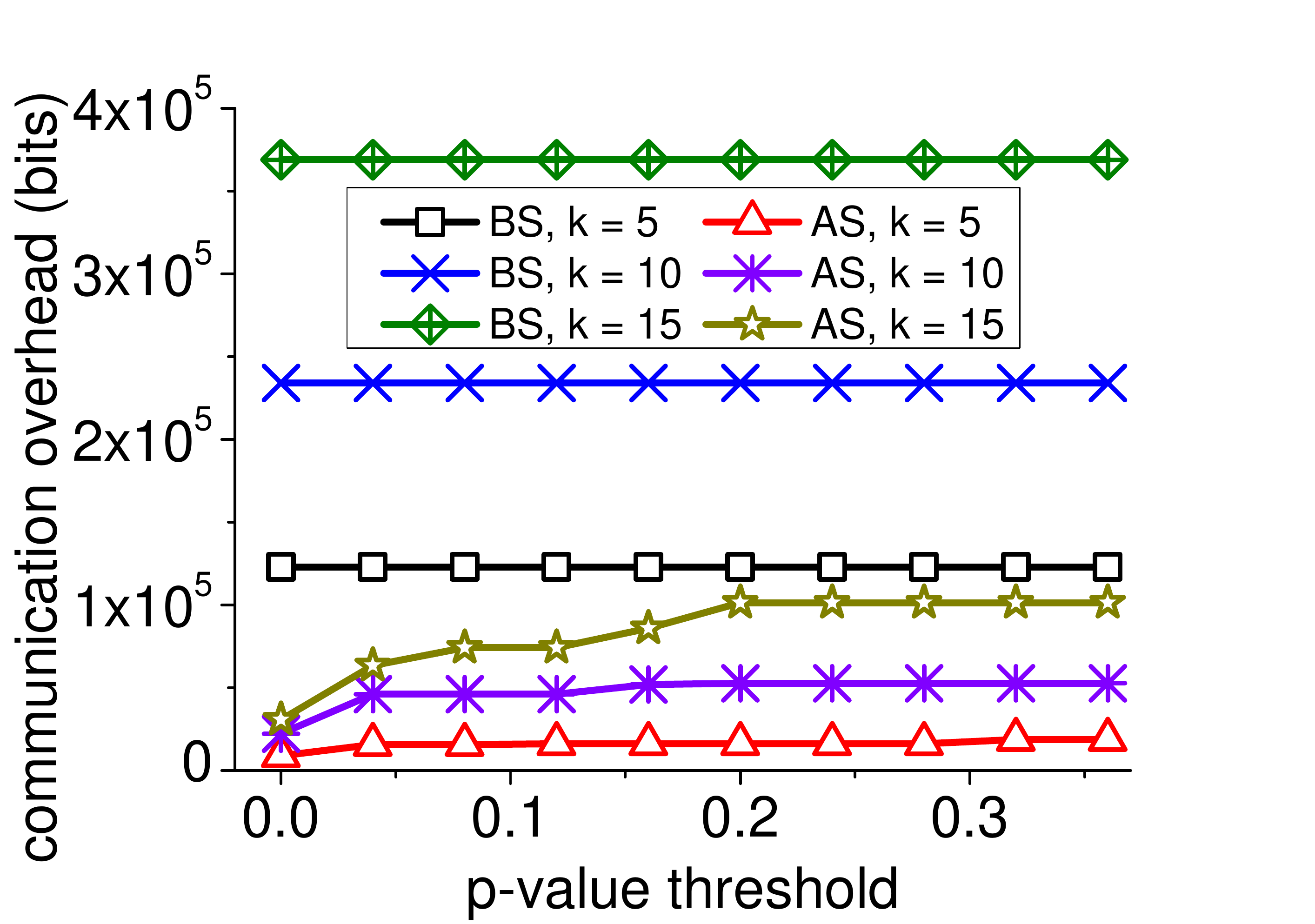}}\hfill
\subfigure[Tag-comp. overhead]{\label{fig:p-tag-comp}
\includegraphics[width=1.8in]{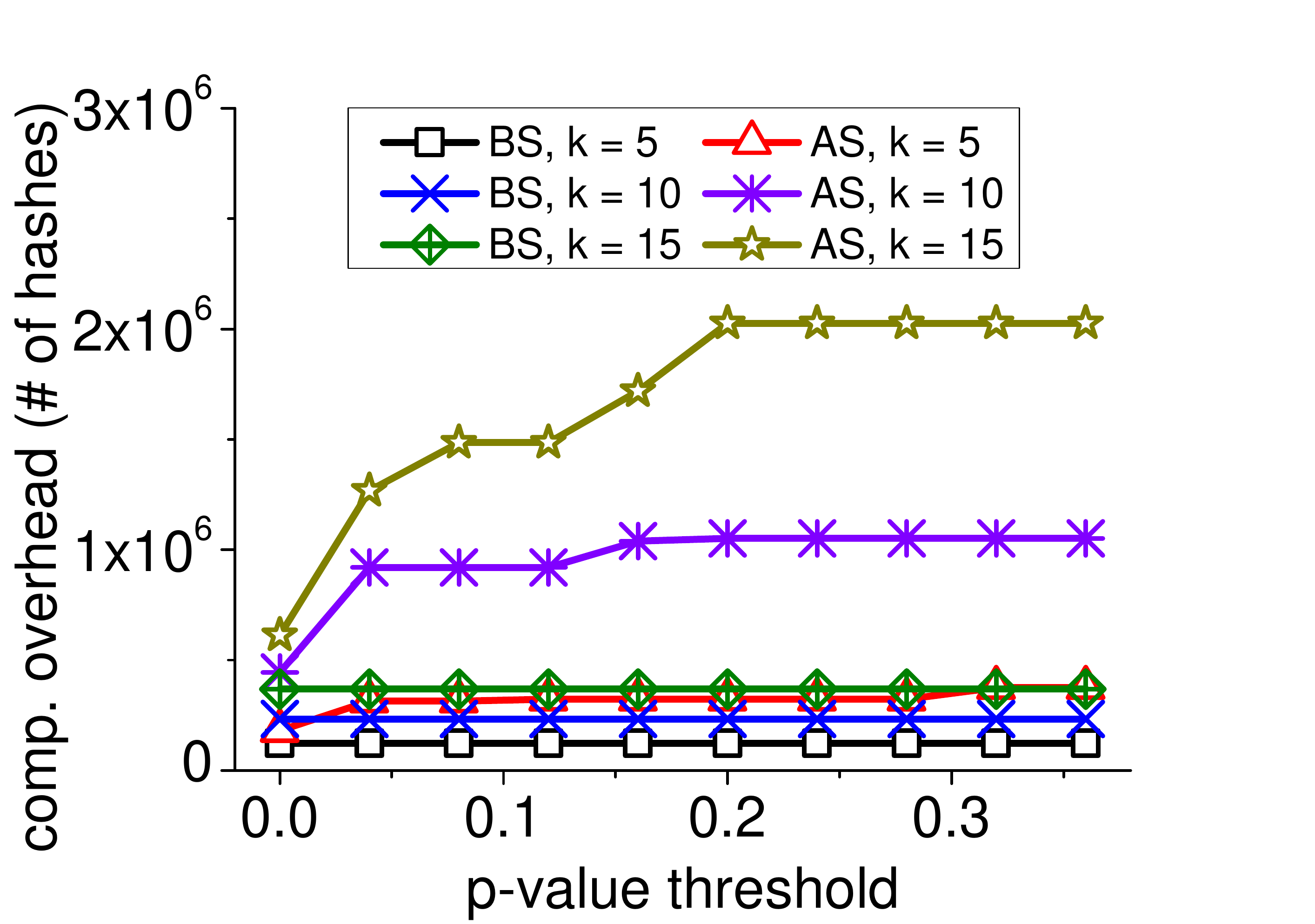}}\hfill
\subfigure[Detector-comm. overhead]{\label{fig:p-detector-comm}
\includegraphics[width=1.8in]{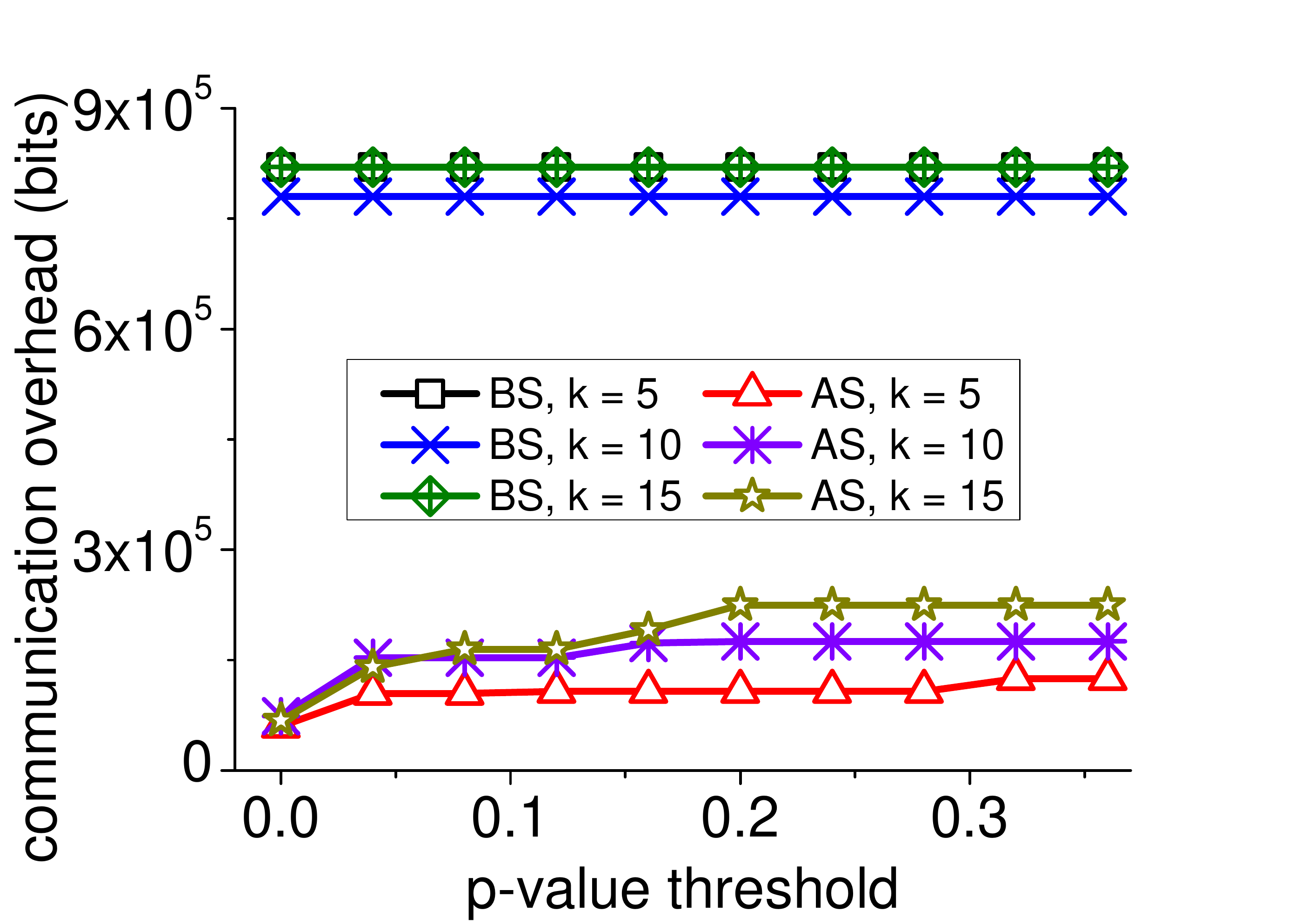}}\hfill
}
\caption{Impact of $p_{\textrm{thre}}$, where BS and AS stand for the basic and advanced schemes, respectively.}  \label{fig:pthreshold}
\end{figure*}

\begin{figure*}[t]
\centerline{
\subfigure[Normalized rank]{\label{fig:k-rank}
\includegraphics[width=1.8in]{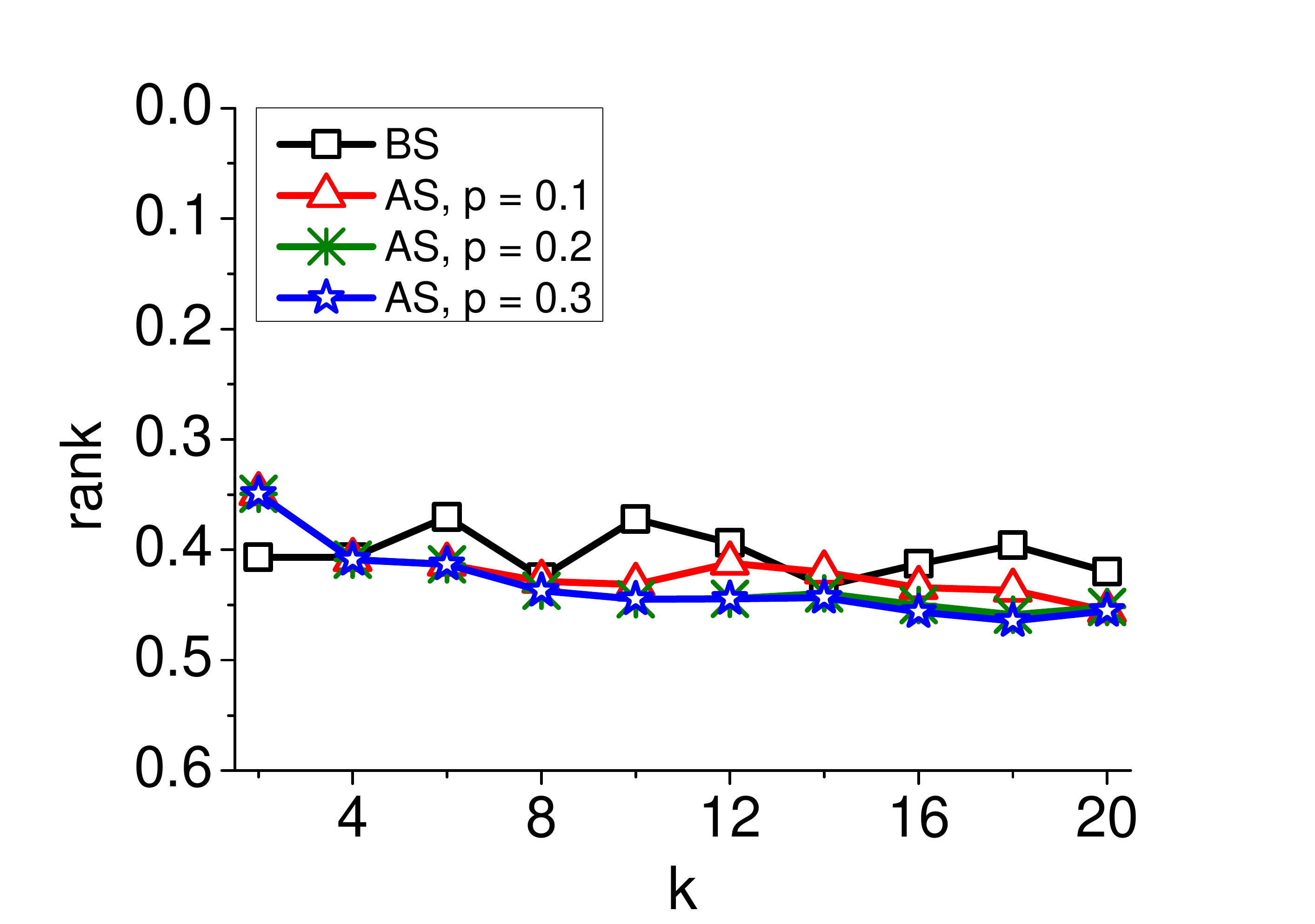}}\hfill
\subfigure[Tag-comm. overhead]{\label{fig:k-tag-comm}
\includegraphics[width=1.8in]{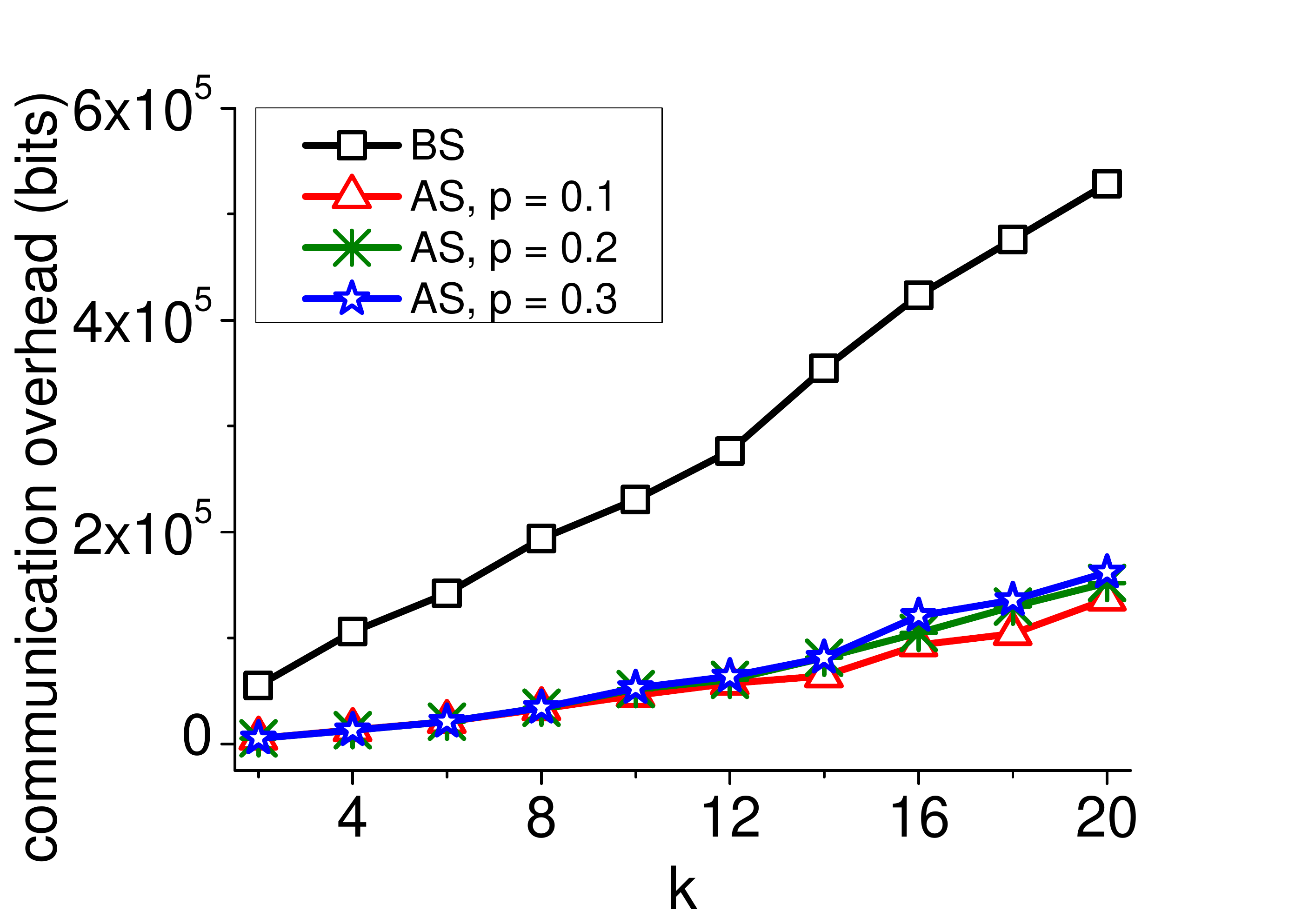}}\hfill
\subfigure[Tag-comp. overhead]{\label{fig:k-tag-comp}
\includegraphics[width=1.8in]{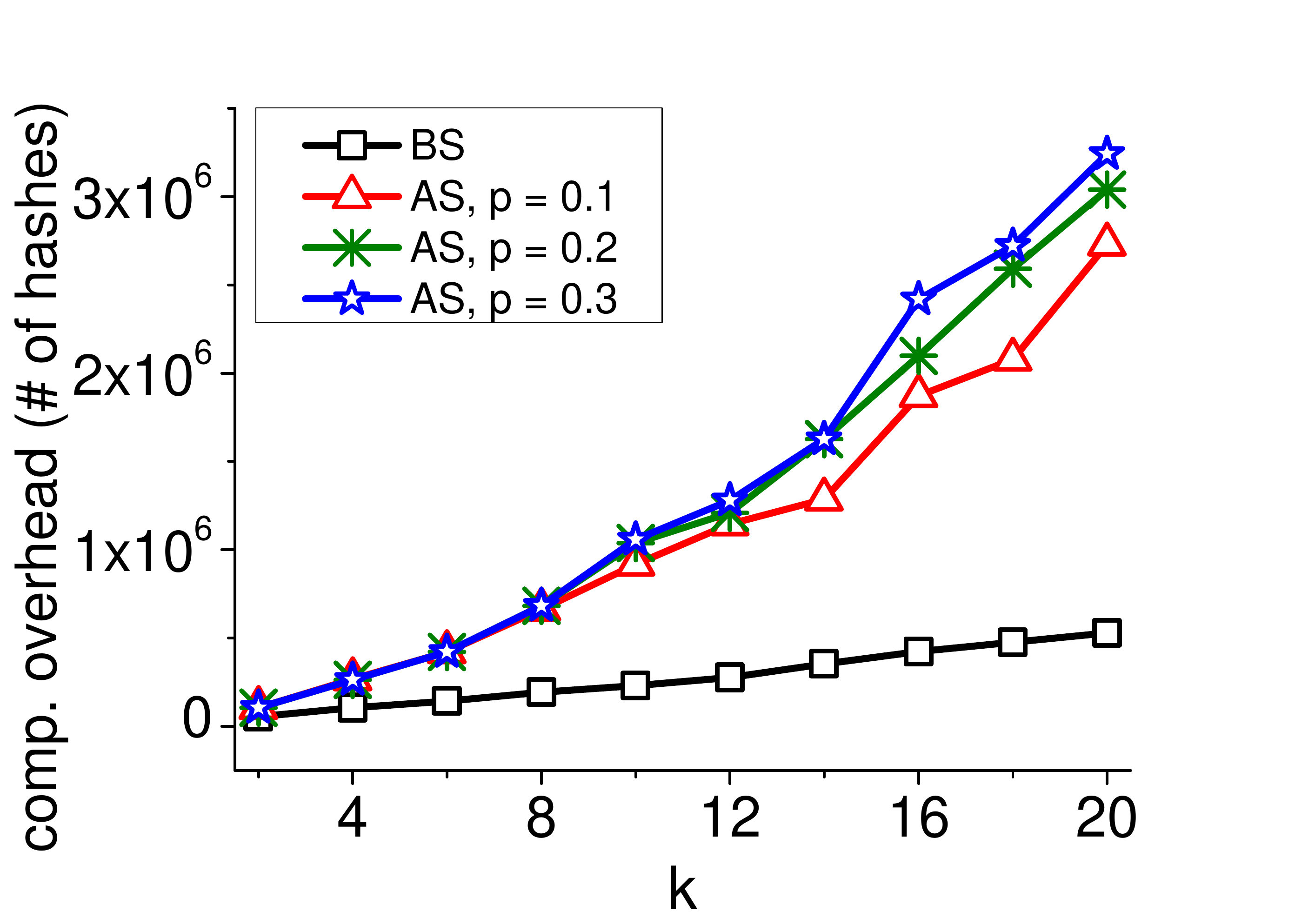}}\hfill
\subfigure[Detector-comm. overhead]{\label{fig:k-detector-comm}
\includegraphics[width=1.8in]{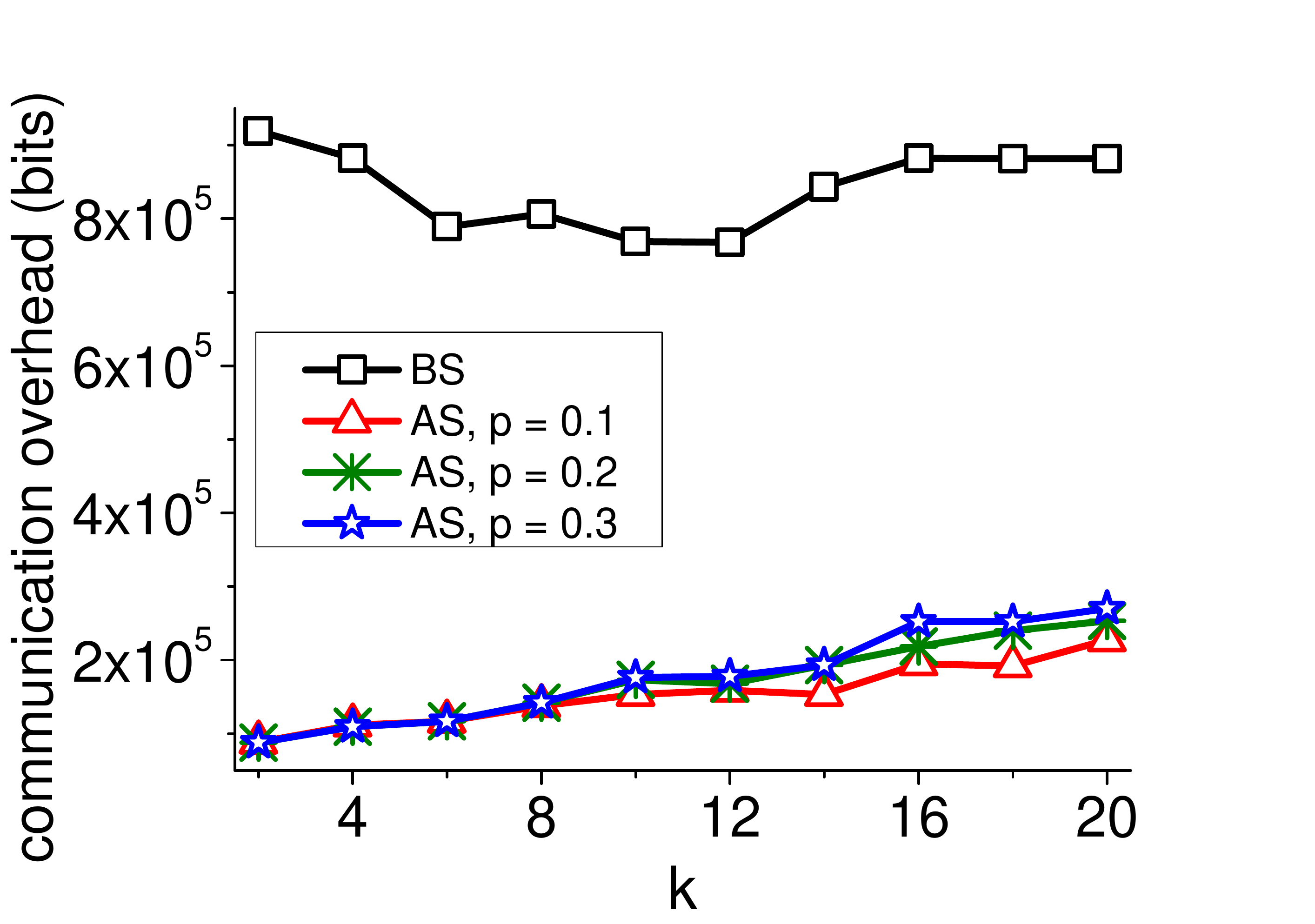}}\hfill
} \caption{Impact of $k$.}  \label{fig:k}
\end{figure*}

%%%%%%%%%%%%%%%%%%%%%%%%%%%%%%%%%%%%%%%%%%%%%%%%%%%%%%%%%%%%%%%
\section{Performance Evaluation}\label{sec:Evaluation}
%%%%%%%%%%%%%%%%%%%%%%%%%%%%%%%%%%%%%%%%%%%%%%%%%%%%%%%%%%%%%%%

In this section, we evaluate the proposed schemes using extensive simulations. We consider a square region with a side length of $2,000$m, in which 10,000 mobile detectors are distributed uniformly, each acting as a dummy tag with probability $q=0.9$. The transmission ranges of mobile detectors and the lost tag are both $50$ m. For our purpose, the simulation code is written in Java, and each data point represents an average of 100 simulation runs each with a different random seed. Table~\ref{tab:Parameter} summarizes our default simulation parameters if not mentioned otherwise.

\begin{table}%[t]%\footnotesize
\small
\renewcommand{\arraystretch}{1.1}
\caption{Default Simulation Settings} \label{tab:Notation} \centering
\begin{tabular}{c|c|l}
\hline \textbf{Para.} & \textbf{Value}&\textbf{Meaning}\\
\hline
$C$& 10000& The number of mobile detectors\\
$q$& 0.9& The probability of acting as dummy tag\\
$f$& 300& The frame length in Frame Slotted ALOHA\\
$k$& 10& The number of hash functions\\
$\omega$& 15& The number of polled positions\\
%$q$& 0.9& prob. of a detector as dummy tag \\
%& 15& The length of tag query \\
%$p_{\textrm{thre}}$& &$p$-value threshold set by owner\\
\hline
\end{tabular}\label{tab:Parameter}
\end{table}

Since the basic and advanced schemes can both offer mobile detectors' strong location privacy and also ensure that the lost object is recoverable almost for sure in all our simulations, our subsequent evaluation focuses on object security, communication overhead, and computation overhead. We assume that the following strategy is adopted by the service provider. On receiving the polling results from all the detectors, the service provider runs the Pearson's chi-squared test as the owner does in the advanced scheme and computes a $p$-value for each detector. The service provider then ranks all the detectors based on their $p$-values. The lower the $p$-value of a detector, the more likely that the lost tag is in his coverage. We then use the relative rank of the detector covering the lost tag to measure the security of the lost object. If the lost tag is covered by multiple detectors, we use the highest rank available. Note that this strategy is a generalization of ranking collectors according to the numbers of bit-one positions discussed in Section~\ref{sec:analysis-basic}, as it additionally considers the possible different numbers of dummy tags around each collector. %Finally, a tag in our results refer to the lost tag and also a dummy tag.

\begin{figure*}[!t]
\centerline{
\subfigure[Normalized rank]{\label{fig:f-rank}
\includegraphics[width=1.8in]{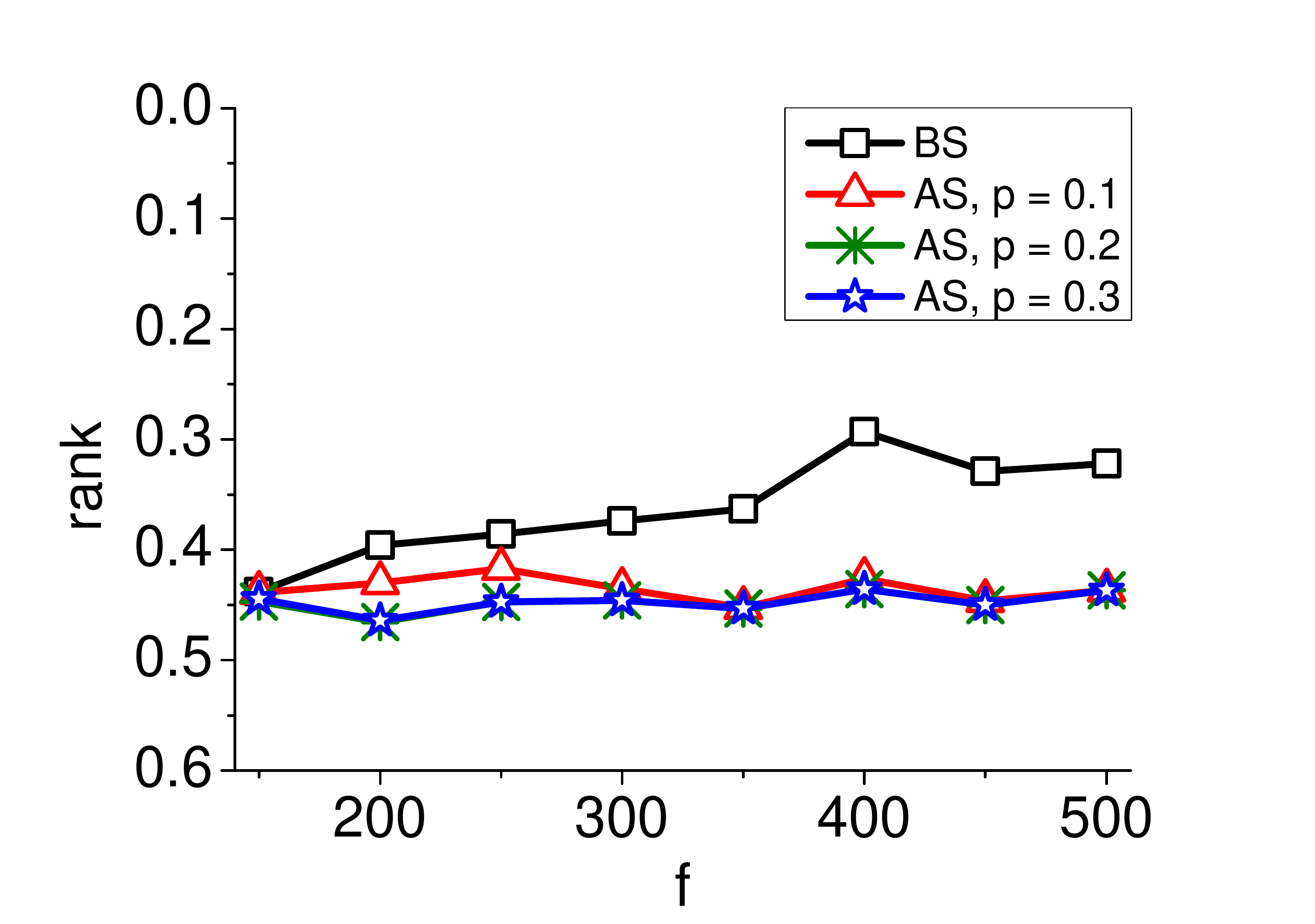}}\hfill
\subfigure[Tag-comm. overhead]{\label{fig:f-tag-comm}
\includegraphics[width=1.8in]{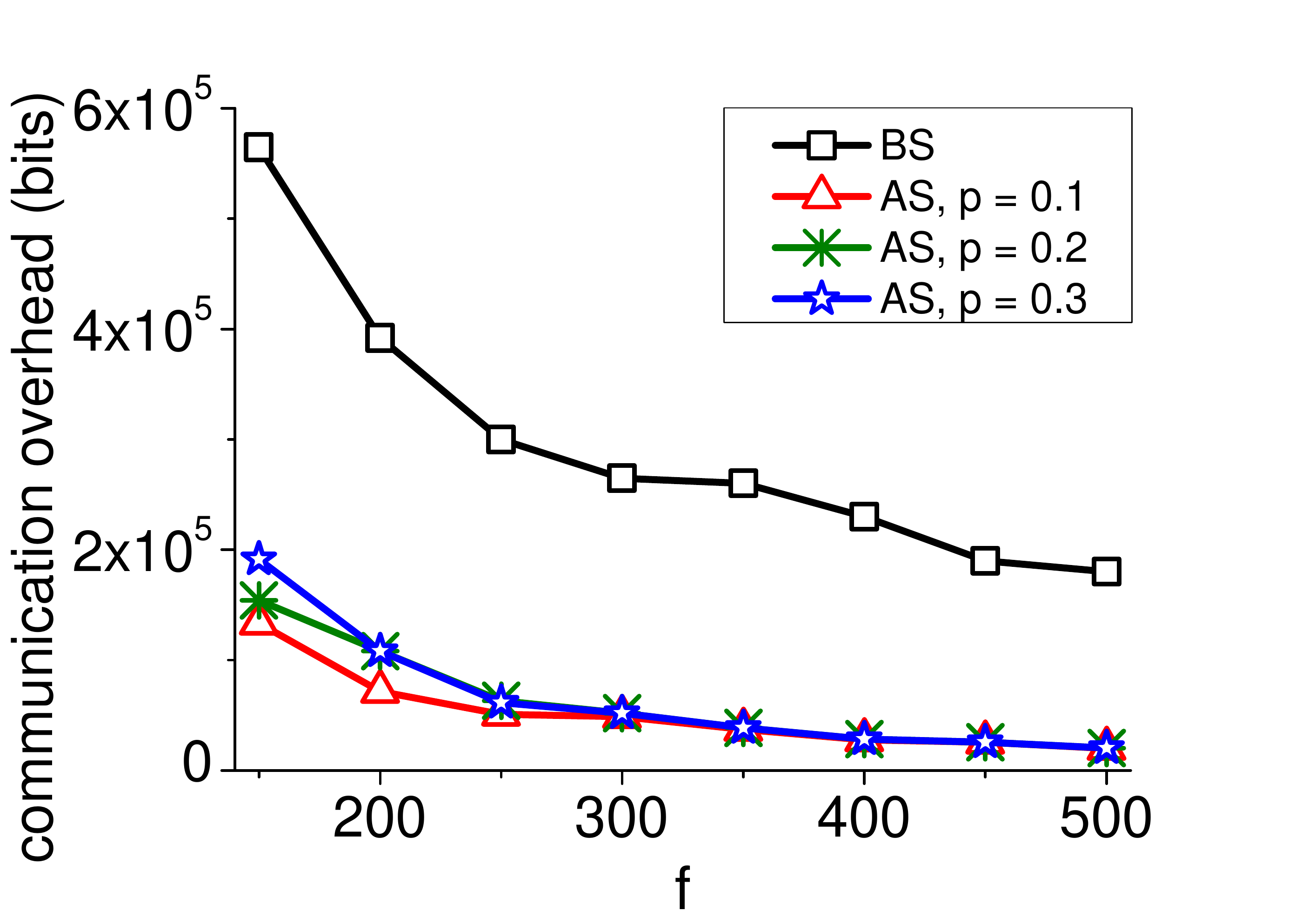}}\hfill
\subfigure[Tag-comp. overhead]{\label{fig:f-tag-comp}
\includegraphics[width=1.8in]{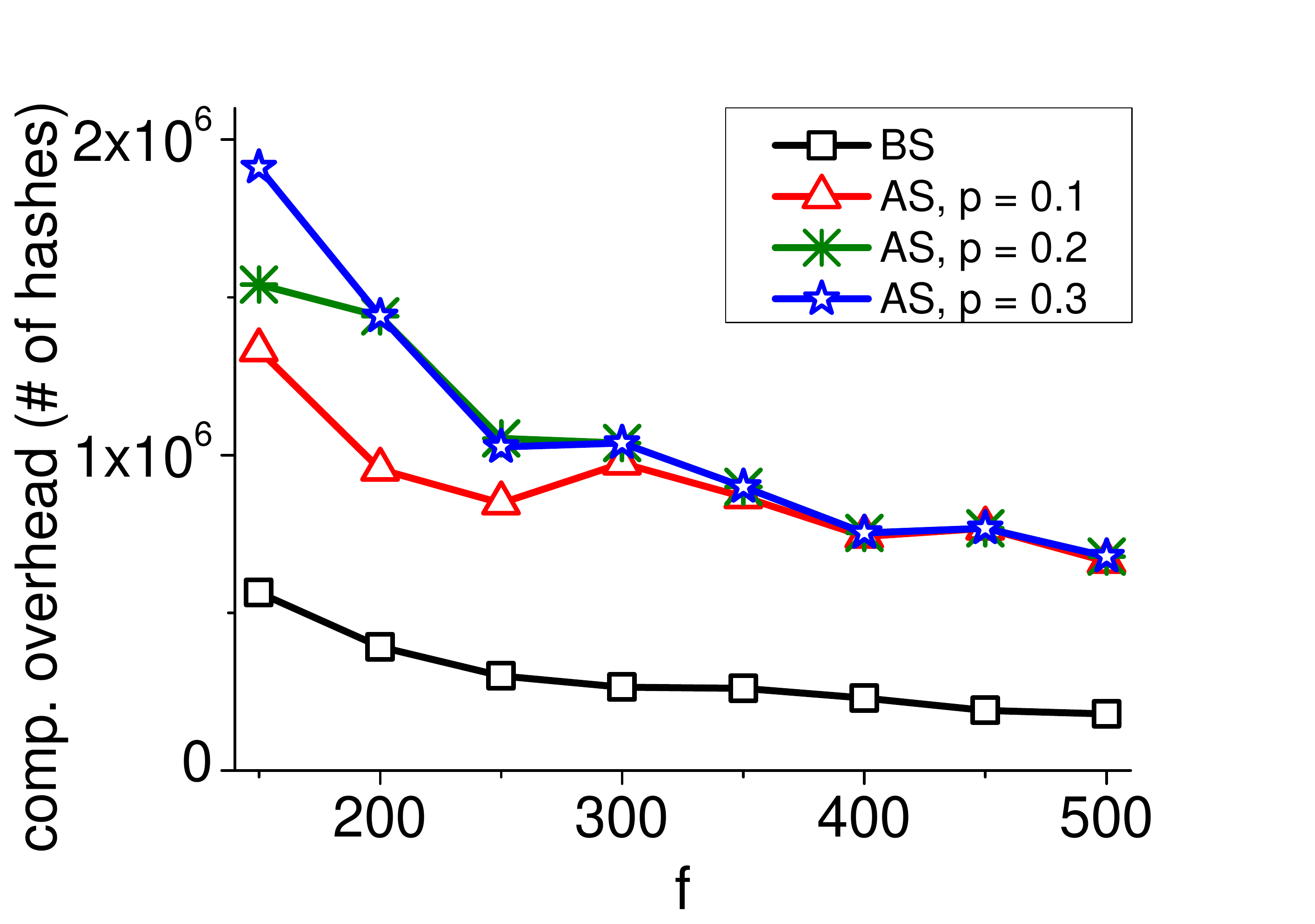}}\hfill
\subfigure[Detector-comm. overhead]{\label{fig:f-detector-comm}
\includegraphics[width=1.8in]{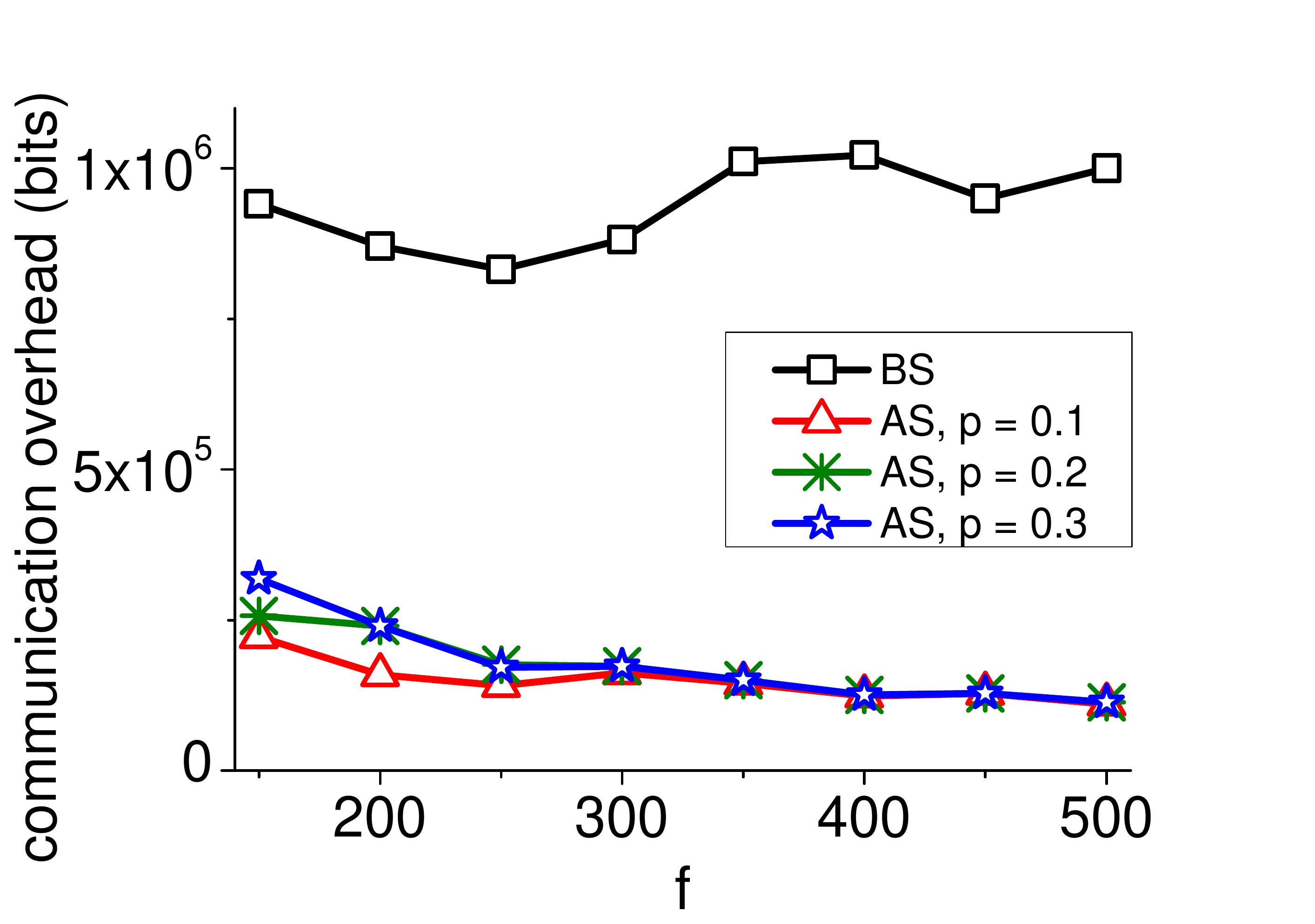}}\hfill
} \caption{Impact of $f$.}  \label{fig:f}
\end{figure*}

\begin{figure*}[!t]
\centerline{
\subfigure[Normalized rank]{\label{fig:w-rank}
\includegraphics[width=1.8in]{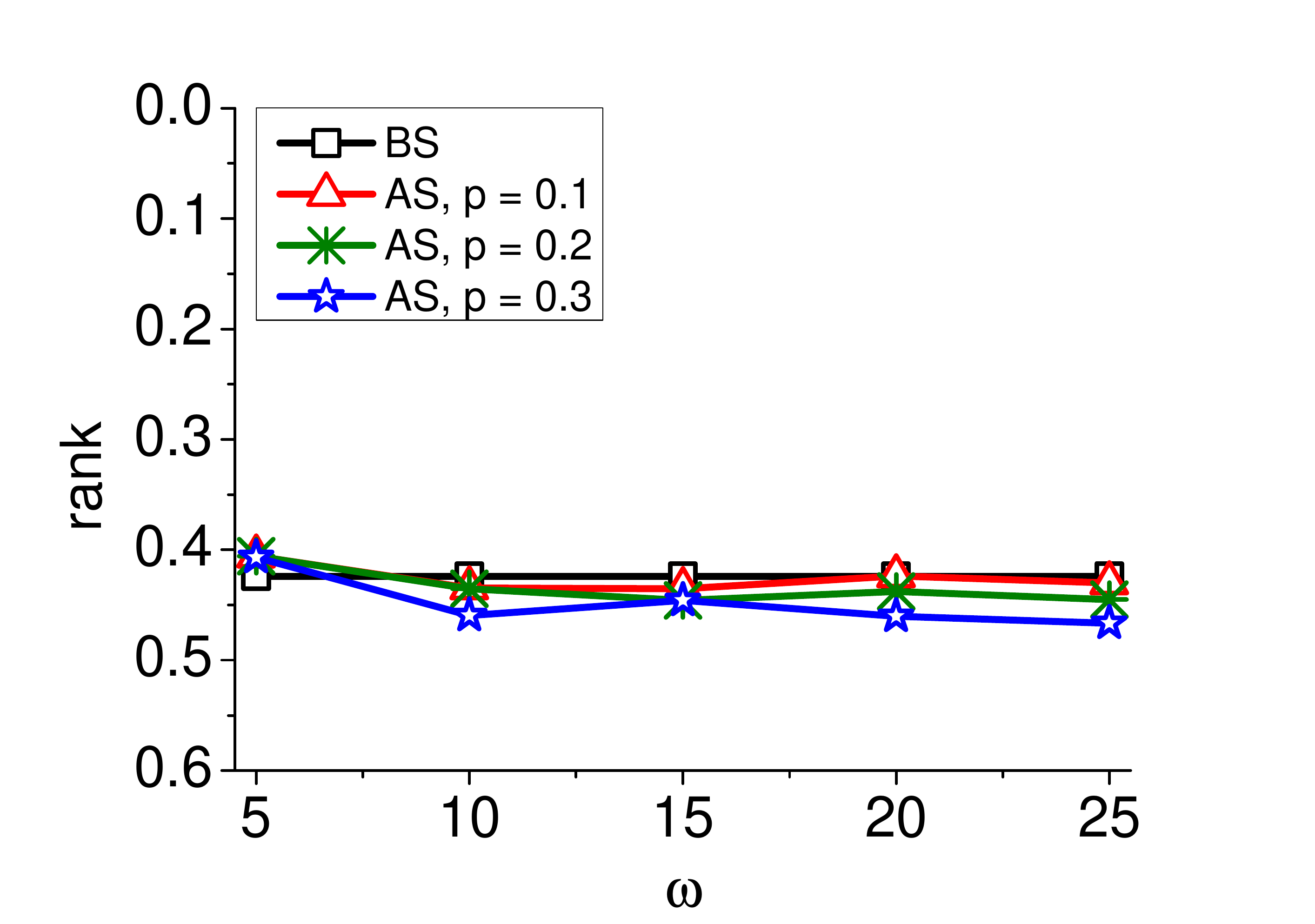}}\hfill
\subfigure[Tag-comm. overhead]{\label{fig:w-tag-comm}
\includegraphics[width=1.8in]{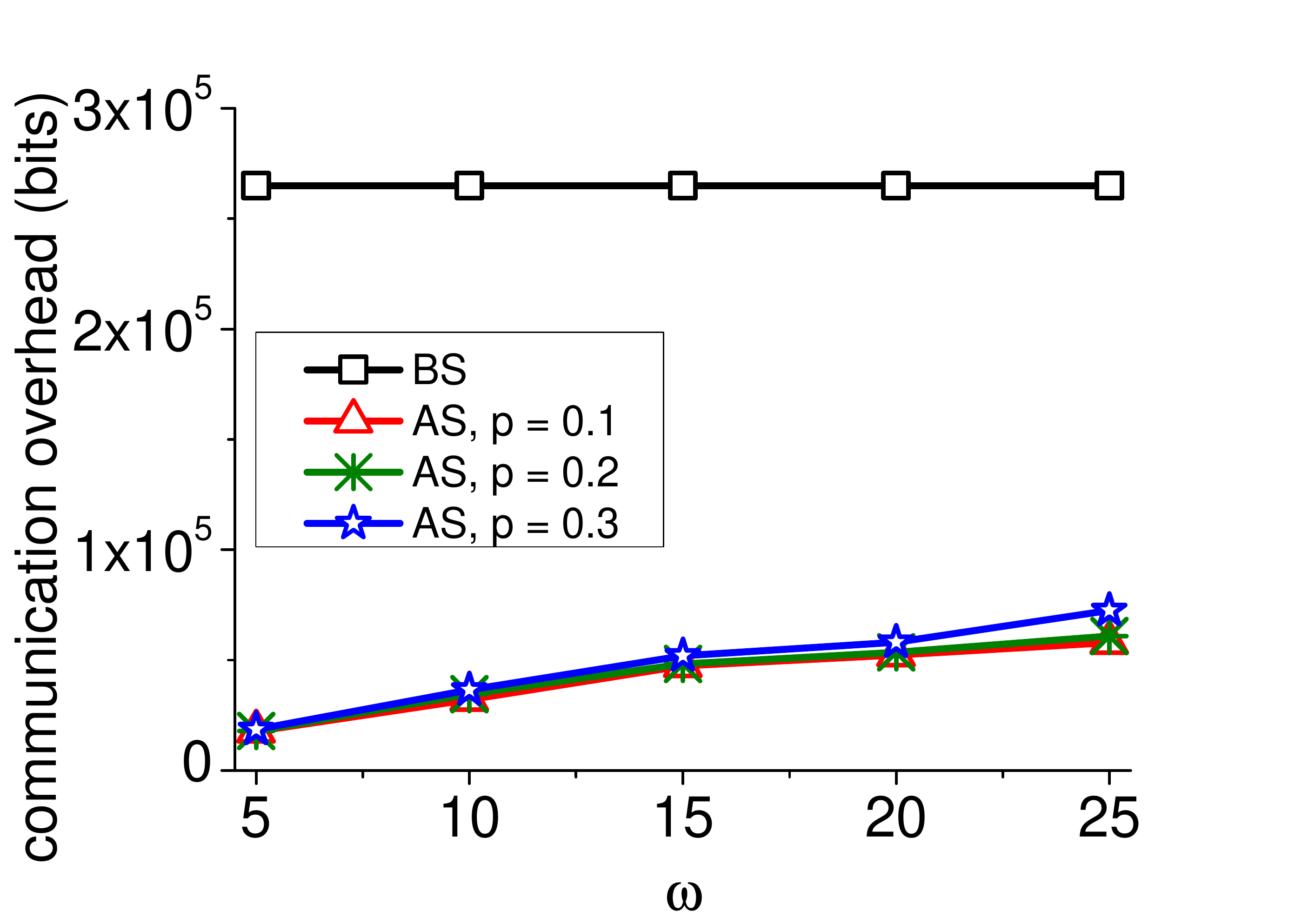}}\hfill
\subfigure[Tag-comp. overhead]{\label{fig:w-tag-comp}
\includegraphics[width=1.8in]{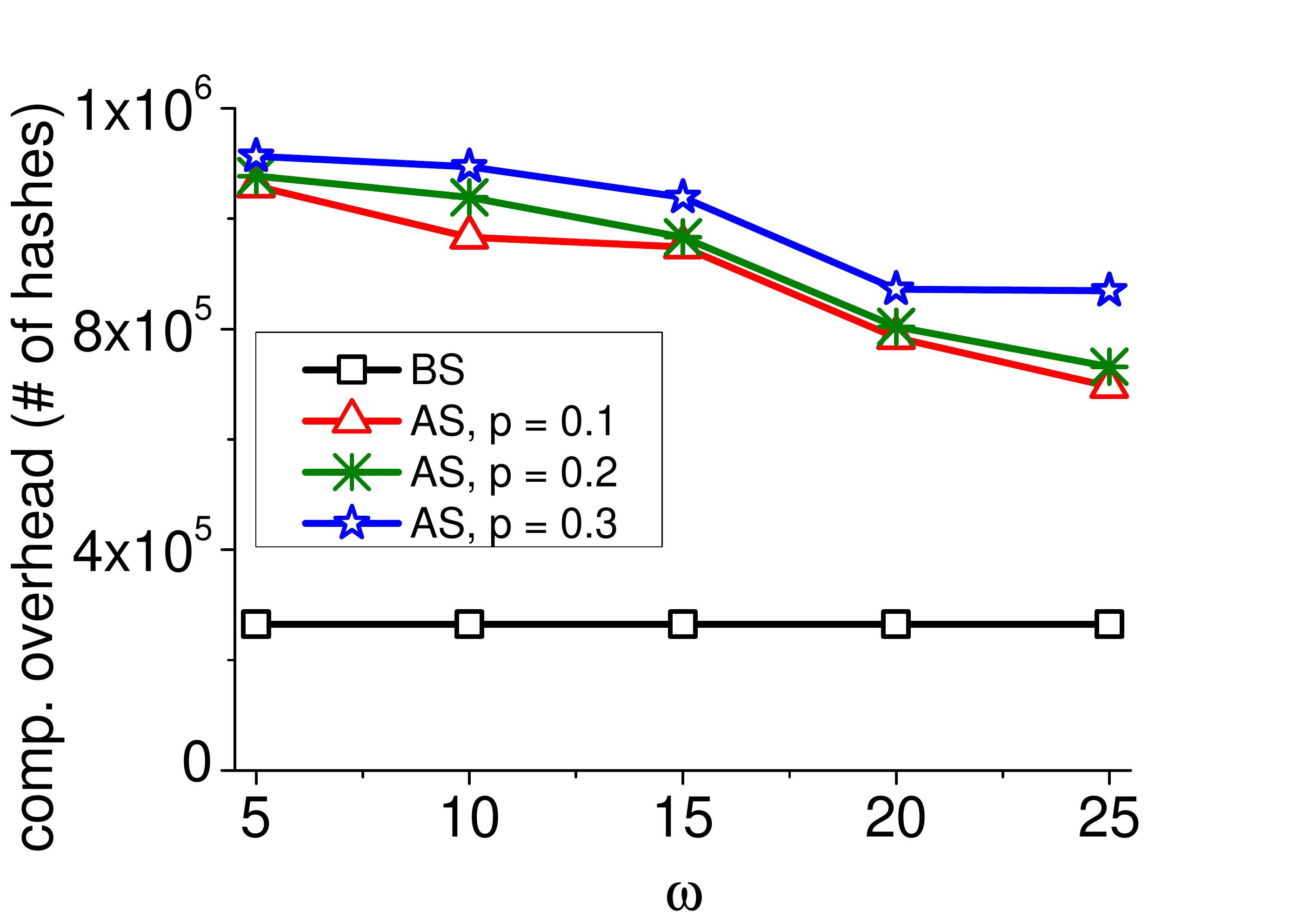}}\hfill
\subfigure[Detector-comm. overhead]{\label{fig:w-detector-comm}
\includegraphics[width=1.8in]{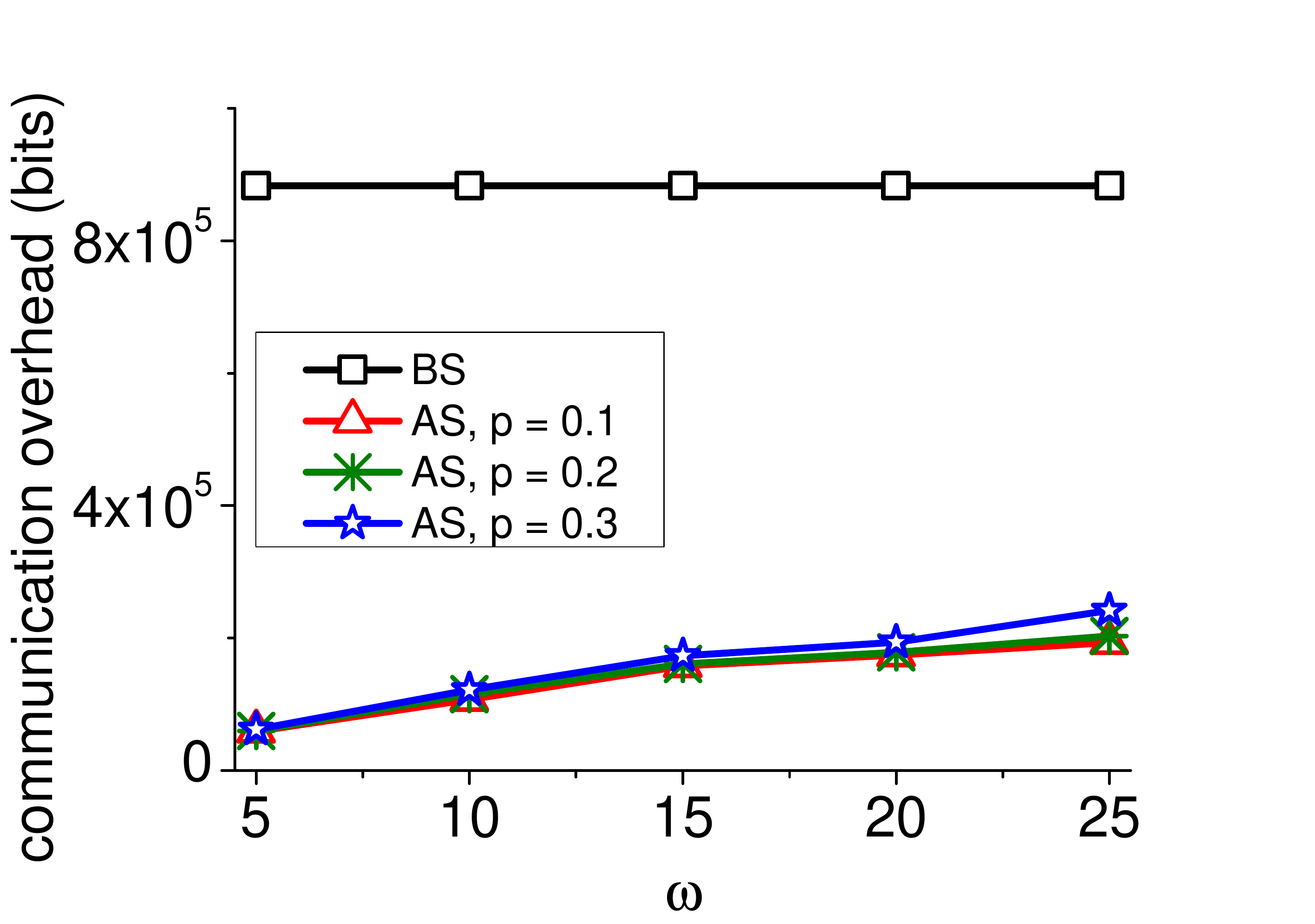}}\hfill
} \caption{Impact of $\omega$.}  \label{fig:w}
\end{figure*}

\vspace{.1in}\noindent \textbf{Impact of $p_{\textrm{thre}}$}. Figs.~\ref{fig:p-rank} to \ref{fig:p-detector-comm} show the object security in terms of the real detector's normalized rank, tag-communication overhead, tag-computation overhead in the number of hash computations performed, and detector-communication overhead of the basic and advanced schemes, respectively. Since the basic scheme is not affected by $p_{\textrm{thre}}$ (the $p$-value threshold), its performance is plotted for reference only. We can see from Fig.~\ref{fig:p-rank} that as $p_{\textrm{thre}}$ increases from 0 to 0.3, the real detector's normalized rank under the advanced scheme increases from around 0.1 to 0.4. This is anticipated, as the higher  $p_{\textrm{thre}}$, the fewer real positions polled in each polling round, the smaller the gap between $p_1$ and $p'_1$, the lower the rank of the real detector, the higher object security, and vice versa. In addition, we can see from Figs.~\ref{fig:p-tag-comm} to \ref{fig:p-detector-comm} that the tag-communication overhead, tag-computation overhead, and  detector-communication overhead of the advanced scheme all increase as $p_{\textrm{thre}}$ increases. The reason is that higher $p_\textrm{thre}$ leads to fewer real positions polled in each round and thus more polling rounds needed to locate the lost object. Moreover, the advanced scheme incurs higher tag-computation overhead than the basic scheme, as the advanced scheme requires more polling rounds than the basic scheme and thus each every tag to perform more hash computations. Finally, Figs.~\ref{fig:p-tag-comm} and \ref{fig:p-detector-comm} show that the advanced scheme incurs lower tag- and detector-communication overhead than the basic scheme. This is of no surprise because  much fewer bits are transmitted from each detector to the service provider in each round under the advanced scheme.

\vspace{.1in}\noindent \textbf{Impact of $k$}. Figs.~\ref{fig:k-rank} to \ref{fig:k-detector-comm} compare the basic and advanced schemes when $k$ (the number of hash functions) varies from 2 to 20. We can see from Fig.~\ref{fig:k-rank} that the real collector's normalized rank fluctuates as $k$ increases under both schemes. The reason is that the increase in $k$ leads to higher $p_1$ for the real detector as well as higher $p'_1$ for fake collectors, which nevertheless has little impact on the gap between $p_1$ and $p'_1$. In addition, Figs.~\ref{fig:k-tag-comm} shows that the tag-communication overhead of both schemes increases with $k$. The reason is that the larger $k$ is, the more slots every tag needs to respond in each polling round, which leads to higher tag-communication overhead. In addition, the advanced scheme incurs much lower communication overhead than the basic scheme, which is expected. Moreover, we can see from Fig.~\ref{fig:k-tag-comp} that the tag-computation overhead of both schemes increases as $k$ increases and that the advanced scheme incurs higher computation overhead. The reason is that the larger $k$ is, the more hash computations each tag needs to perform in each polling round. In addition, since we generally have $\gamma<k$ in the advanced scheme, it requires more rounds to locate the lost tag, while every tag needs to perform $k$ hash computations in each round.

\vspace{.1in}\noindent \textbf{Impacts of $f$}. Figs.~\ref{fig:f-rank} to \ref{fig:f-detector-comm} show the object security in terms of the real detector's normalized rank, tag-communication overhead, tag-computation overhead in the number of hash computations performed, and detector-communication overhead of the basic and advanced schemes, respectively. Similar to $k$, $f$ has very limited impact on the normalized rank of the real detector. In addition, we can see from Fig.~\ref{fig:f-tag-comm} and Fig.~\ref{fig:f-tag-comp} that the tag-communication and tag-computation overhead of both schemes decrease as $f$ increases. The reason is that the larger $f$, the fewer polling rounds needed to locate the lost tag, the lower tag-communication and tag-computation overhead for both schemes, and vice versa. In addition, the advanced scheme incurs lower tag-communication overhead but higher tag-computation overhead. Moreover, we can see from Fig.~\ref{fig:f-detector-comm} that the detector-communication overhead of the advanced scheme decreases as $f$ increases. The reason is that in each polling rounds, each detector needs to transmit a $\omega$-bit vector which is not affected by $f$. Fewer polling rounds thus lead to lower detector-communication overhead. In contrast, the detector-communication overhead of the basic scheme remains stable as $f$ increases. The reason is that the detector-communication overhead of the basic scheme is the product of the number of polling rounds and the frame length. Since the increase in $f$ leads to the decrease in the number of polling rounds, the detector-communication overhead of the basic scheme is relatively stable. 

\vspace{.1in}\noindent \textbf{Impacts of $\omega$}. Figs.~\ref{fig:w-rank} to \ref{fig:w-detector-comm} show the impact of $\omega$ on the advanced scheme, where the performance of the basic scheme is plotted for reference only.  We can see from Fig.~\ref{fig:w-rank} that $\omega$ has very limited impact on the object security. In addition, we can see from Figs.~\ref{fig:w-tag-comm} to \ref{fig:w-detector-comm} that the tag-communication and detector communication overhead both increase and the tag-computation overhead decreases as $\omega$ increases.

\section{Conclusion}\label{sec:Conclusion}
%%%%%%%%%%%%%%%%%%%%%%%%%%%%%%%%%%%%%%%%%%%%%%%%%%%%%%%%%%%%%%%
This paper presented the design, analysis, and evaluation of SecureFind, the first secure and privacy-preserving crowdsourced object-finding system. In particular, we first introduced a basic scheme which provides strong object security at the cost of system efficiency, and then presented an advanced scheme to strike a good balance between object security and system efficiency. Detailed simulations confirmed that SecureFind can enable very fast and efficient object finding while ensuring the security of the lost object and also the location privacy of the mobile users participating in object finding.

There are still many open challenges to tackle. For example, in our current design, all the mobile detectors in the target area specified by the object owner need to participate in object finding. Since some of them may have overlapping coverage, there may be significant room for reducing the communication and computation overhead. One possible solution is to let the service provider select the minimum number of mobile detectors that can jointly cover the target area. This solution, however, requires the service provider to know more accurate locations of mobile detectors. Such tradeoff between system efficiency and location privacy deserves careful investigation. In addition, our current design assumes that mobile detectors are honest-but-curious. There may be dishonest mobile detectors who report fake search results to earn reward without actually performing the object search. How to catch and then punish such dishonest mobile detectors is nontrivial and may conflict with the location-privacy requirement of mobile detectors. We hope that this paper can stimulate further interest in crowdsourced object finding and other exciting mobile crowdsourcing applications.

\bibliographystyle{IEEETran}

\end{document}